\documentclass[prd,preprint,showpacs]{revtex4-1}
\usepackage{amsmath,latexsym,amssymb,amsfonts,color}
\usepackage{graphicx}

\newcommand{\beq}{\begin{equation}}
\newcommand{\eeq}{\end{equation}}
\newcommand{\bea}{\begin{eqnarray}}
\newcommand{\eea}{\end{eqnarray}}
\newcommand\beano{\begin{eqnarray*}}
\newcommand\eeano{\end{eqnarray*}}

\newcommand{\al}{\alpha}
\newcommand{\be}{\beta}
\newcommand{\ga}{\gamma}

\newcommand{\del}{\delta}
\newcommand{\eps}{\epsilon}
\newcommand{\la}{\lambda}

\newcommand{\LL}{{\cal L}}

\newcommand{\csta}{{X}}
\newcommand{\cstb}{{Y}}
\newcommand{\cstc}{{Z}}
\newcommand\ltwid{\mathrel{
 \raise.3ex\hbox{$<$\kern-.75em\lower1ex\hbox{$\sim$}}}}

\begin{document}

\preprint{UdeM-GPP-TH-14-238}

\title{Tunneling decay of false kinks}
\author{\'Eric Dupuis$^{a}$}
\email{eric.dupuis.1@umontreal.ca}
\author{Yan Gobeil$^{a}$}
\email{yan.gobeil@umontreal.ca}
\author{Richard MacKenzie$^{a}$}
\email{richard.mackenzie@umontreal.ca}
\author{Luc Marleau$^{b}$}
\email{luc.marleau@phy.ulaval.ca}
\author{M.~B.~Paranjape$^{a}$}
\email{paranj@lps.umontreal.ca}
\author{Yvan Ung$^{a}$}
\email{yvan.ung@umontreal.ca}
\affiliation{$^a$Groupe de physique des particules, Universit\'{e} de Montr\'{e}al, C.~P.~6128, Succursale Centre-ville, Montr\'eal, QC, Canada, H3C 3J7}
\affiliation{$^b$D\'epartement de physique, de g\'enie physique et d'optique, Universit\'e Laval, Qu\'ebec, QC, Canada G1K 7P4}

\begin{abstract}
We consider the decay of ``false kinks,'' that is, kinks formed in a scalar field theory with a pair of degenerate symmetry-breaking false vacua in 1+1 dimensions.  The true vacuum is symmetric. A second scalar field and a peculiar potential are added in order for the kink to be classically stable. We find an expression for the decay rate of a false kink. As with any tunneling event, the rate is proportional to $\exp(-S_E)$ where $S_E$ is the Euclidean action of the bounce describing the tunneling event. This factor varies wildly depending on the parameters of the model. Of interest is the fact that for certain parameters $S_E$ can get arbitrarily small, implying that the kink is only barely stable. Thus, while the false vacuum itself may be very long-lived, the presence of kinks can give rise to rapid vacuum decay.
\end{abstract}

\maketitle

\newpage

\section{Introduction \label{sec-intro}}

Kinks are topological solitons in 1+1-dimensional field theories with a real scalar field $\phi$ and spontaneously-broken discrete symmetry $\phi\to-\phi$. The potential has degenerate vacua at field values $\phi=\pm v$; the kink interpolates between the two. The same model in higher dimensions gives rise to extended objects: linelike defects in 2+1 dimensions, domain walls in 3+1 dimensions, and so on. 

\begin{figure}[hbt]
\begin{center}
\includegraphics[width=3.in]{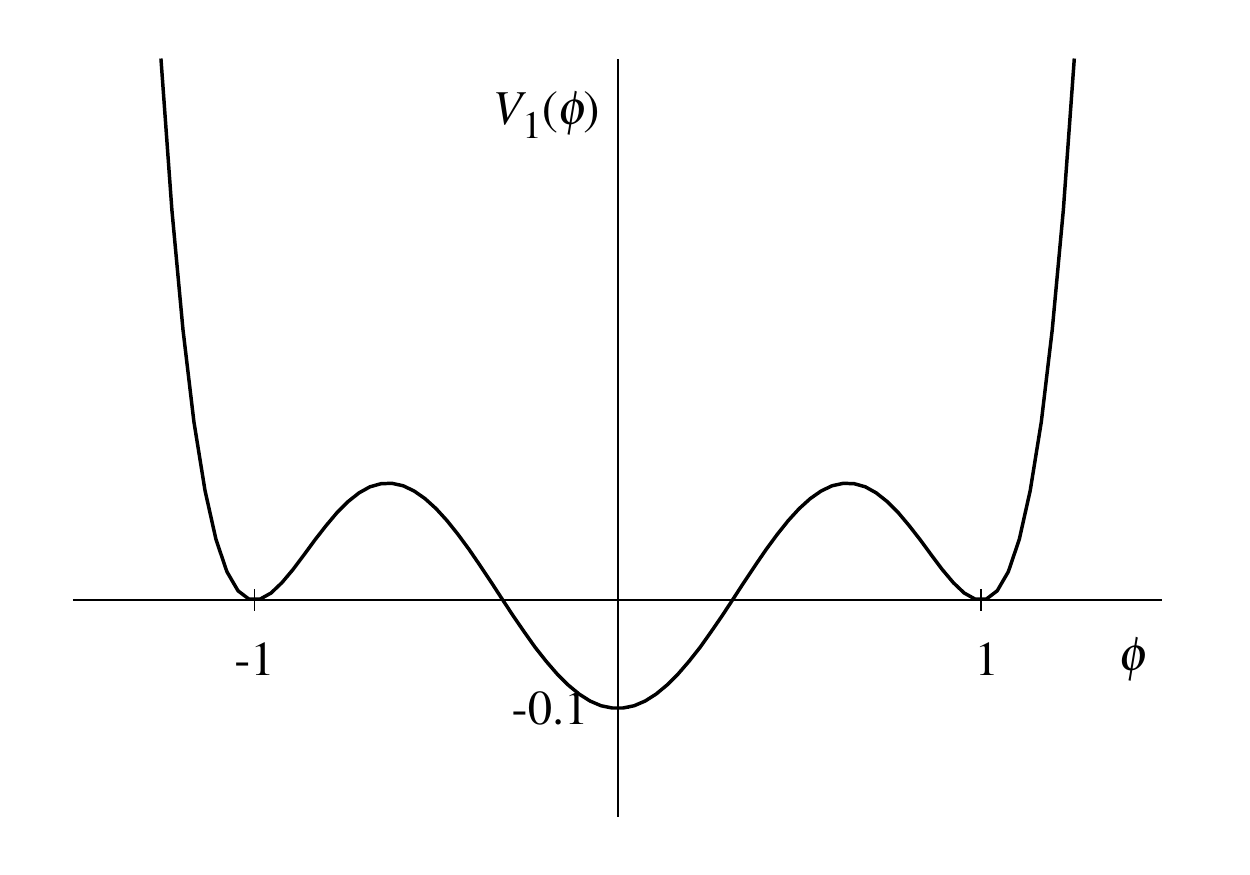}
\end{center}
\caption{Example potential (\ref{eq-pot1}) with symmetric true vacuum and symmetry-breaking false vacua. Here $\la=v=1,\ \del=0.1$.
}
\label{fig1}
\end{figure}

We are interested in models which ``demote" the vacua at $\pm v$ to {\em false vacuum} status, there being a lower-energy {\em true vacuum} at $\phi=0$. A specific example is
\beq
\label{eq-lagrangian0}
\LL_\phi = \frac{1}{2}(\partial_\mu\phi)^2 - V_1(\phi)
\eeq
where (see Fig.~\ref{fig1})
\beq
\label{eq-pot1}
V_1(\phi) = \la (\phi^2 - \del v^2) (\phi^2 - v^2)^2
\eeq
with $0<\del<1$. In the true vacuum the symmetry is of course restored. The motivation for studying such a model is that the presence of topological defects can have a dramatic effect on the quantum mechanical stability of the false vacuum. Using cosmological language for convenience, if the universe is in a false vacuum (without kinks) throughout space, it will decay through quantum tunneling \cite{Coleman:1977a}, yet the decay rate per unit volume can be exceedingly small, to the point where the observable universe could be trapped in a false vacuum for times exceeding the age of the universe. Such a scenario is invoked in certain models of fundamental physics; see for instance \cite{Kachru:2003a}. If so, the presence of topological defects can have a dramatic effect on the decay rate. (It should be noted that if they exist, topological defects will necessarily be formed during a phase transition, so their presence is not merely a possibility; it is a certainty \cite{Kibble:1976a}.)  This situation has been examined previously for magnetic monopoles \cite{Kumar:2010a}, vortices in 2+1 dimensions \cite{Lee:2013a}, and cosmic strings \cite{Lee:2013b}.

Since there is a unique true vacuum in the class of model we are considering, there are no classically stable nontrivial static solutions. For instance, a kinklike configuration interpolating between the two false vacua would bifurcate into two halves, one interpolating between $\phi=-v$ and $\phi=0$ and the other between $\phi=0$ and $\phi=v$. The potential energy density being lower between the two halves of the kink than in the exterior region, the kinks would essentially repel each other and fly off to spatial infinity at speeds approaching that of light, leaving an ever-expanding region of true vacuum.

Thus we must consider a more complicated model if we wish to study the decay of classically stable kinks formed in a symmetry-breaking false vacua. The paper is outlined as follows. In Section 2, we introduce a generalized model and argue that it does have classically stable kinks. In Section 3, we find numerical solutions for a variety of parameters. These solutions, although classically stable, will tunnel to the true vacuum, a process analyzed in Section 4. The comparison between kink-mediated vacuum decay and ordinary vacuum decay is analyzed in Section 5. We then present conclusions and suggestions for further work.

\section{A model with classically stable kinks \label{sec-model}}

As argued above, classically stable kinks do not exist in the simplest model in which the vacuum structure is as outlined above (two symmetry-breaking false vacua, symmetric true vacuum). One way to obtain stable kinks is to add a second scalar field to the model with an unusual potential. Consider the following model, which is to be viewed more as an example having the desired vacuum structure and stable kinks rather than as a realistic model for a particular physical system.
\[
\LL = \frac{1}{2}\left( (\partial_\mu\phi)^2 + (\partial_\mu\chi)^2\right) - V(\phi,\chi)
\]
where
\[
V(\phi,\chi) = \la_1 (\phi^2 - \del_1 {v_1}^2) (\phi^2 - {v_1}^2)^2
+ \frac{\la_2}{\phi^2+\ga v_1^2} \left[ (\chi^2-v_2)^2 - (\del_2 / 4)(\chi-2v_2)(\chi+v_2)^2 \right]
\]
The model is symmetric under $\phi\to-\phi$. The potential has seven parameters, two of which can be eliminated by rescaling the fields and $x$ to dimensionless variables. Doing so, we can rewrite the (dimensionless) Lagrangian as follows, having chosen the constants in such a way as to simplify the equations of motion:
\beq
\label{eq-lagrangian1}
\LL = \frac{1}{2}(\partial_\mu\phi)^2 + \frac{\al}{2 \be}(\partial_\mu \chi)^2 - V(\phi,\chi)
\eeq
where
\beq
\label{eq-pot2}
V(\phi,\chi) = (\phi^2 - \del_1) (\phi^2 - 1)^2
+ \frac{\al}{\phi^2+\ga} \left[ (\chi^2-1)^2 - \frac{\del_2}{4}(\chi-2)(\chi+1)^2 \right].
\eeq
There are now five parameters, $\al,\be,\del_1,\del_2,\ga$, which we take to be positive; furthermore, we suppose $0<\del_1<1$ and $0<\del_2<16/3$ in order for the potential to have the properties we desire. Although it entails a loss of generality, in what follows we will assume $\al=\be=1$ for simplicity.

\begin{figure}[hbt]
\begin{center}
\includegraphics[width=3.in]{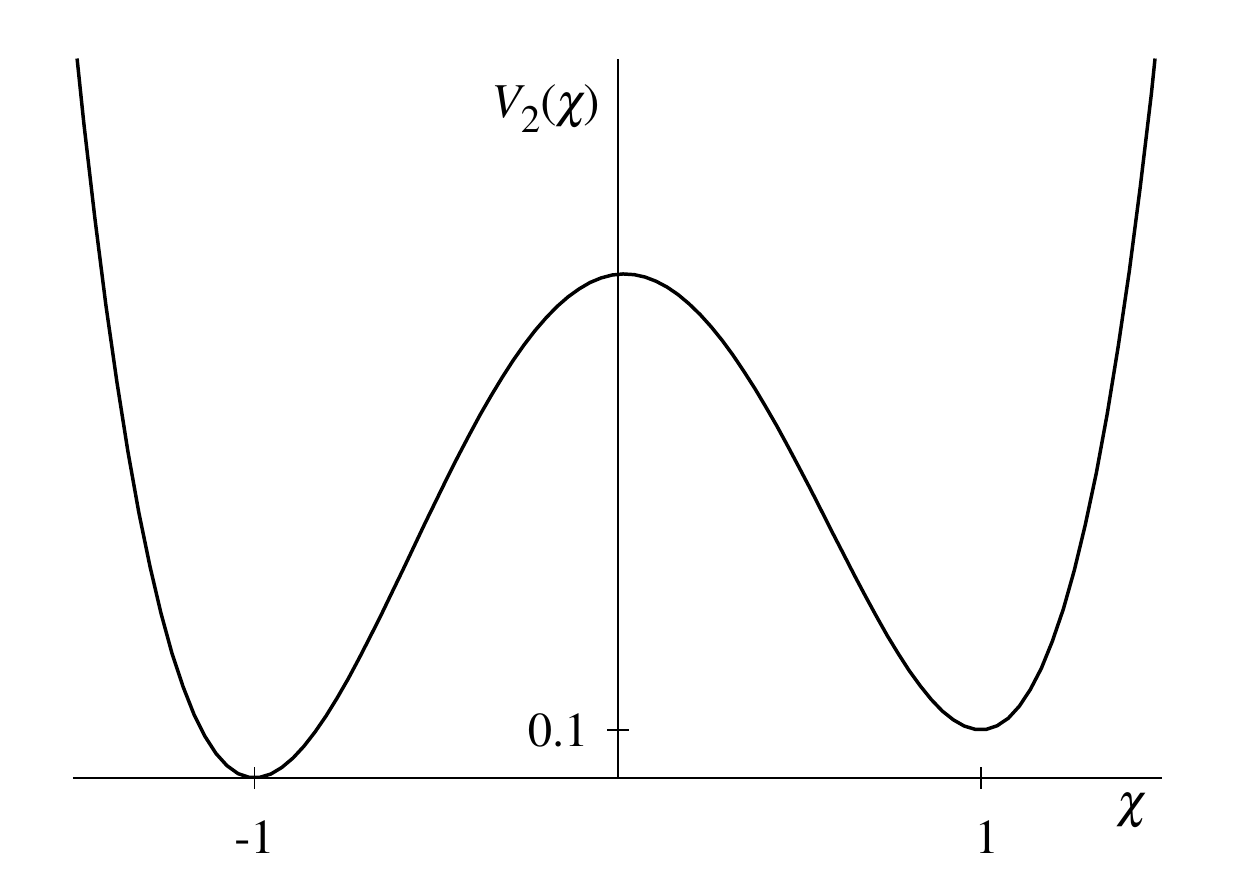}
\end{center}
\caption{Last factor of the potential \eqref{eq-pot2}. Here $\del_2=0.1$.
}
\label{fig2}
\end{figure}

The potential is the sum of two terms. The first term depends on $\phi$ only and is, apart from rescaling, the potential \eqref{eq-pot1} of the original model (see Fig.~\ref{fig1}). The second term is a product of two factors. The second of these (in square parentheses, written $V_2(\chi)$ below; see Fig.~\ref{fig2}) depends on $\chi$ only. With $\del_2$ in the above-mentioned range, $V_2$ has two minima: a global minimum, of zero energy density, at $\chi=-1$ and a local minimum, of energy density $\delta_2$, at $\chi=+1$. The first factor can be viewed as a modulating function which varies the ``strength" of the second factor depending on the value of $\phi$. In particular, if $\phi=0$ (its true vacuum), the modulating factor is maximal, so that if $\chi$ passes from one minimum to the other, the cost in potential energy where $\chi\simeq0$ will be large.

\begin{figure}[hbt]
\begin{center}
\includegraphics[width=3.5in]{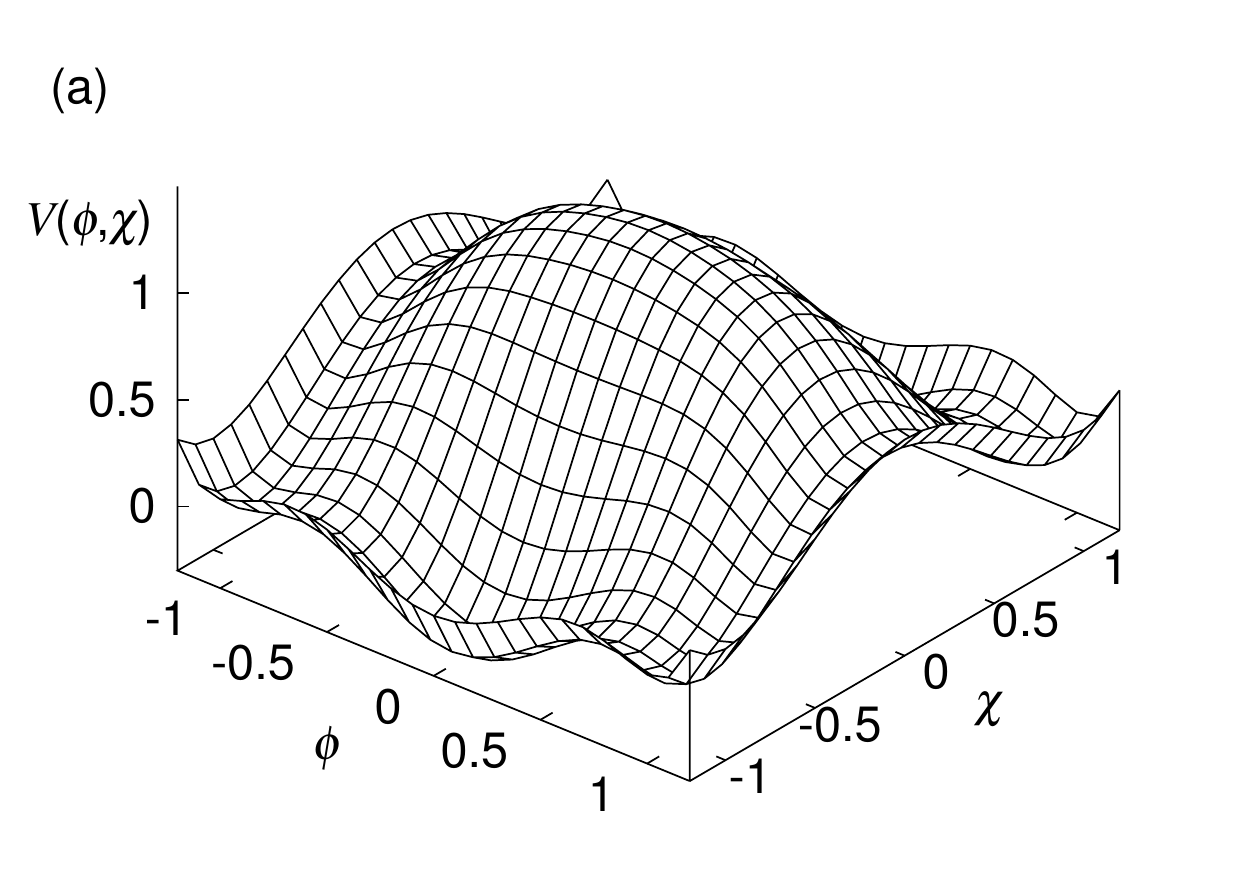}
\includegraphics[width=2.8in]{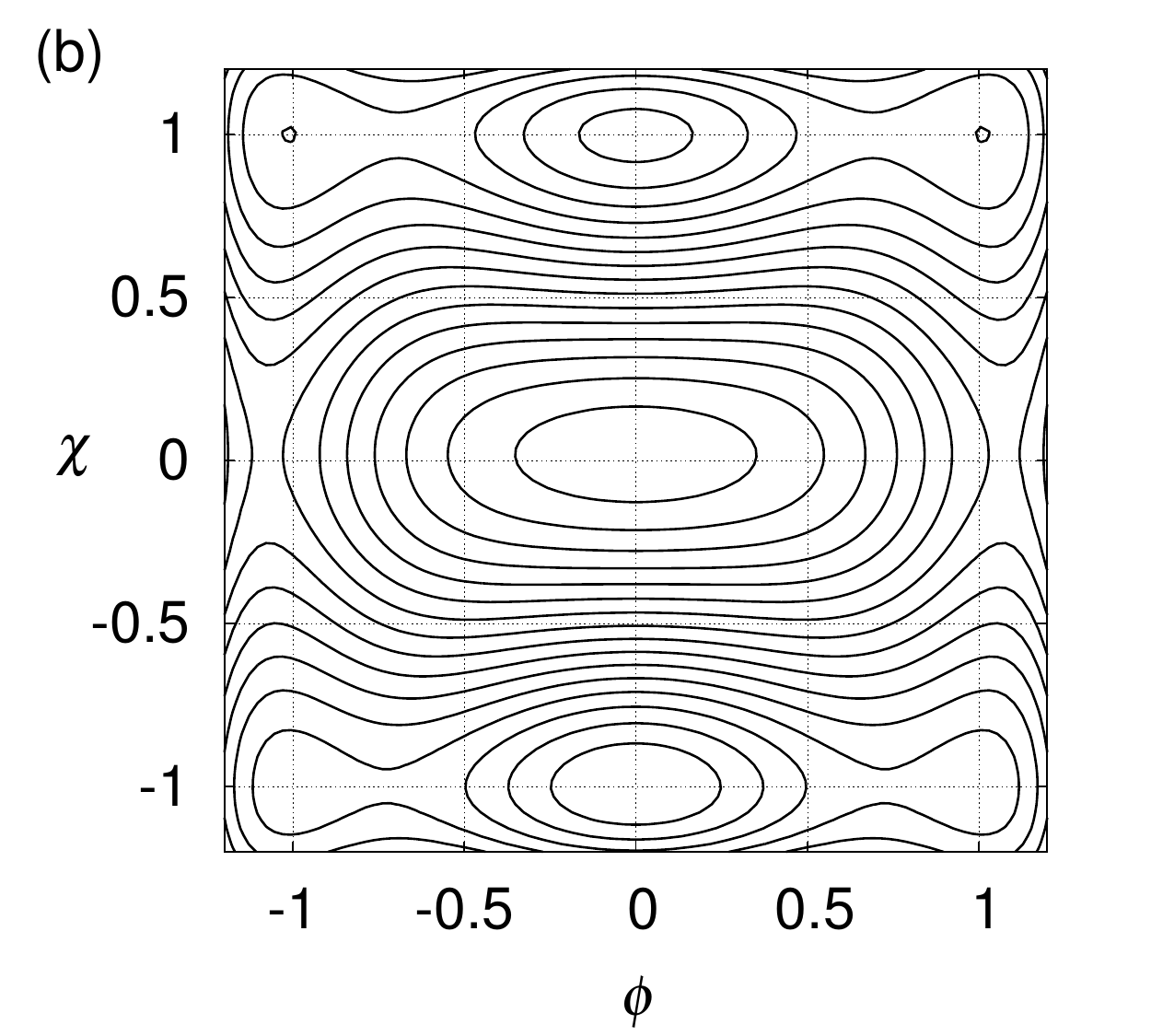}
\end{center}
\caption{(a) 3-dimensional plot and (b) contour plot of $V(\phi,\chi)$ for 
%$(\al,\del_1,\del_2,\ga)=(1,0.3,0.1,0.7)$.
$(\del_1,\del_2,\ga)=(0.3,0.1,0.7)$.
(These parameter values do not produce stable kink solutions; they are used here simply to illustrate the form of the potential.) The true vacuum is at $(\phi,\chi)=(0,-1)$; there are false vacua at $(\phi,\chi)=(\pm1,-1)$. There are also extrema along the line $\chi=1$, the nature of which depend on the values of the parameters, as described in the text. (In the figure, they are local minima -- false vacua -- at $(\phi,\chi)=(0,1)$ and very near $(\pm1,1)$.) The potential energy density is maximal, and can be very large, near $(\phi,\chi)=(0,0)$.
}
\label{fig3}
\end{figure}

The relevant features of the potential, illustrated in Fig.~\ref{fig3}, are as follows. The true vacuum of the model is $(\phi,\chi)=(0,-1)$; its energy density is $V(0,-1)=-\del_1$.  There are false vacua at $(\phi,\chi)=(\pm1,-1)$, of energy density $V(\pm1,-1)=0$. There is also a maximum in the vicinity of $(\phi,\chi)=(0,0)$ which can be quite pronounced if $\ga\ll 1$. Finally, there are possible extrema at $\chi=1$ which are less important but not entirely irrelevant to us, so it is worth examining briefly the potential there:
\beq
V(\phi,1) = (\phi^2 - \del_1) (\phi^2 - 1)^2 + \frac{\del_2}{\phi^2+\ga}.
\eeq
This is the sum of the potential depicted in Fig.~\ref{fig1} and a Lorentzian function. It is of interest to ask if $V(\phi,1)$ can be negative, since as we will see it gives rise to an instability of the soliton (though not the type of instability which is of primary interest to us). If the Lorentzian is broad and sufficiently large in amplitude, it will raise the potential near $\phi=0$ so that it is nowhere negative (for $\chi=1$). However if the Lorentzian is narrow, it can raise the potential at $\phi=0$ to a positive value while leaving negative regions on either side. Thus there are two cases. In the first case $V(\phi,1)$ is minimized at $\phi=0$ so that $(\phi,\chi)=(0,1)$ is a local minimum of the potential. This occurs if
\beq
\ga>\frac{\del_1}{1+2\del_1}.
\label{cond1}
\eeq
In this case the minimum of $V(\phi,1)$ is negative if
\beq
\del_2<\del_1\ga.
\label{cond2}
\eeq
If \eqref{cond1} is not satisfied, $V(\phi,1)$ is minimized at a pair of nonzero values of $\phi$ straddling $\phi=0$. These points are local minima (false vacua) of the potential while $(\phi,\chi)=(0,1)$ is a saddle point.
%The analog of \eqref{cond2} is calculable but it is very long and we will not present it here. In fact, we will suggest below that $\del_2$ is a more important parameter than $\del_1$ and $\ga$; in due course we will restrict ourselves to $\del_1=\ga=0.01$, so \eqref{cond1} is indeed satisfied.
Now the minimum of $V(\phi,1)$ is negative if the following surprisingly unwieldy condition is satisfied:
\bea
\left(
\frac{1}{32}\right.&&\left.-\frac{\del_1}{16}-\frac{7{\del_1}^2}{64}+\frac{9{\del_1}^3}{64}
-\frac{27{\del_1}^4}{512}
\right)
+\ga\left( \frac{1}{16}-\frac{13\del_1}{32}+\frac{13{\del_1}^2}{64}-\frac{9{\del_1}^3}{128} \right)
-\ga^2\left( \frac{7}{64}+\frac{13\del_1}{64}+\frac{{\del_1}^2}{256} \right)
\nonumber\\
&&-\ga^3\left( \frac{9}{64}+\frac{9\del_1}{128} \right)   -\frac{27\ga^4}{512}
+\left\{
\left(-\frac{1}{64}+\frac{3\del_1}{128}-\frac{11{\del_1}^2}{256}+\frac{9{\del_1}^3}{512} \right)
-\ga\left( \frac{3}{128}+\frac{5\del_1}{128}-\frac{5{\del_1}^2}{512} \right)\right.
\nonumber\\
&&\left.-\ga^2\left( \frac{11}{256}+\frac{5\del_1}{512} \right)   -\frac{9\ga^3}{512}
\right\}
\sqrt{(2-3(\ga-\del_1))^2 + 16(2\ga\del_1+\ga-\del_1)}
+ \del_2<0.
\label{cond3}
\eea

\begin{figure}[hbt]
\begin{center}
\includegraphics[width=4.in]{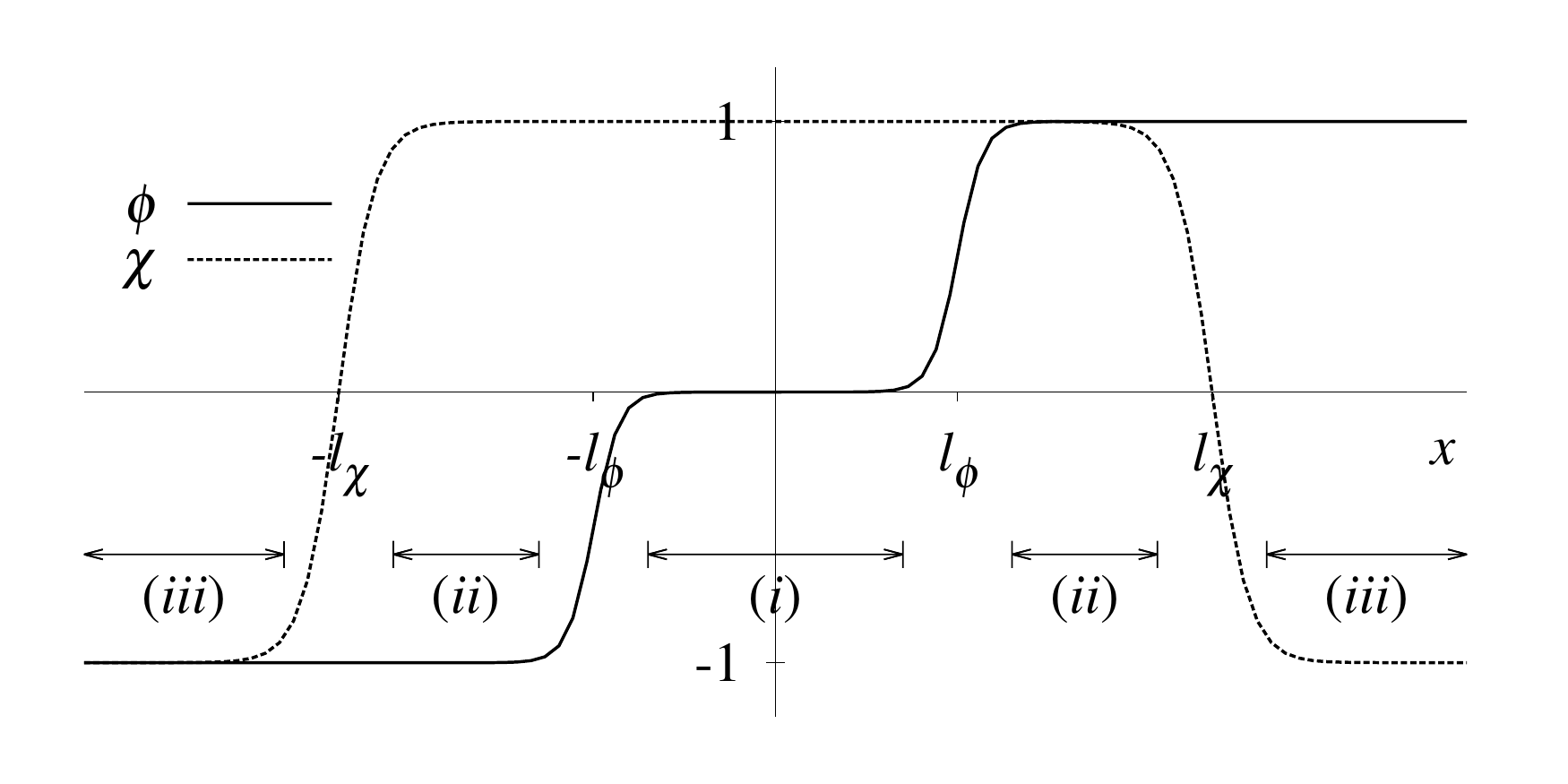}
\end{center}
\caption{Configuration to illustrate intuitively the existence of stable solitons.
}
\label{fig4}
\end{figure}

We can understand qualitatively why there could be stable solitons interpolating between the two false vacua $(\phi,\chi)=(\pm1,-1)$. Consider a configuration of the form illustrated in Fig.~\ref{fig4}, where $\phi$ has half-kinks at $\pm l_\phi$, one interpolating between $\phi=-v$ and $\phi=0$ and the other between $\phi=0$ and $\phi=v$, these being ``enveloped" by a kink-antikink of $\chi$ at $\pm l_\chi$, where $l_\chi > l_\phi$. (To simplify the discussion, we will call all of these objects kinks.) There are five regions where the fields are approximately constant, two pairs of which are related by symmetry; these regions are denoted (i) (between the two $\phi$ kinks), (ii) (the regions between the $\phi$ and $\chi$ kinks), and (iii) (exterior to the $\chi$ kinks). In these regions the energy density comes entirely from the potential energy and is easily evaluated: 
\[
%V_{(i)} = -\del_1+\frac{\al}{\ga}\del_2,\qquad V_{(ii)} = \frac{\al}{1+\ga}\del_2,\qquad V_{(iii)} = 0.\qquad 
V_{(i)} = -\del_1+\frac{1}{\ga}\del_2,\qquad V_{(ii)} = \frac{1}{1+\ga}\del_2,\qquad V_{(iii)} = 0.\qquad 
\]

%The energy difference between the two sides of a kink represents a force which, if the configuration were to evolve dynamically, would cause the kink to accelerate towards the zone of higher potential energy. Since $V_{(iii)}<V_{(ii)}$, the two $\chi$ kinks would move towards the origin. The situation with the $\phi$ kinks is less clear because, depending on the parameters of the model, $V_{(ii)}$ could be larger or smaller than $V_{(i)}$.

It is easy to see that the configuration depicted in Fig.~\ref{fig4} cannot be a solution. Consider a family of such configurations parameterized by $l_\phi,\ l_\chi$, where the positions but not the shapes of the transitions vary. For the configuration to be a solution, its energy must be stationary as a function of $l_\phi,\ l_\chi$. For small displacements $\Delta l_\phi,\ \Delta l_\chi$, the variation in energy is
\[
\Delta E = 2(V_{(i)} - V_{(ii)}) \Delta l_\phi + 2(V_{(ii)} - V_{(iii)}) \Delta l_\chi.
\]
Since $V_{(iii)}<V_{(ii)}$, the energy is {\em not\/} stationary in $l_\chi$; indeed, if left to evolve dynamically, the two $\chi$ kinks would move towards the origin to reduce the static energy. Similarly, since $V_{(iii)}<V_{(i)}$ it is energetically advantageous for region (i) to collapse to zero, converting false vacuum to true.

This variational argument is not powerful enough in itself to determine if a configuration such as that depicted in Fig.~\ref{fig4} would evolve into a stable one, because as soon as the kinks overlap they interact and the energy is no longer a straightforward function of $l_\phi,\ l_\chi$. If the configuration can be deformed to an unstable one without encountering an energy barrier along the way, it will not evolve into a stable one.

One obvious way to go from Fig.~\ref{fig4} to an unstable configuration would be to deform $\chi$ to a constant, $\chi(x)=-1$ (after which the $\phi$ kinks will fly apart, as described above). This could be done by moving the $\chi$ kinks towards one another so that they annihilate, or alternatively by deforming the value of $\chi$ between the kinks from $+1$ to $-1$. In either case, the fields pass through the large potential energy barrier at $(\phi,\chi)\simeq(0,0)$, so there can be a large energy barrier preventing either deformation from occurring.

Of course, this argument is merely suggestive of the existence of stable solitons. In the absence of a convincing analytic argument for their existence, we must resort to numerically solving the equations of motion. This is described in the following section.

\section{Numerical solutions \label{sec-numerical}}

The static equations of motion that follow from \eqref{eq-lagrangian1} (with $\al=\be=1$) are:
\bea
%with \al, \be...
%\phi'' - 2\phi(\phi^2-1)(3\phi^2-1-2\del_1) + \frac{2\al\phi}{(\phi^2+\ga)^2} \left[ (\chi^2-1)^2 - \frac{\del_2}{4}(\chi-2)(\chi+1)^2 \right] &=& 0\\
%\chi'' - \frac{\be}{(\phi^2+\ga)}(\chi^2-1)(4\chi-3\del_2/4) &=& 0
%without...
\label{phieqn}
\phi'' - 2\phi(\phi^2-1)(3\phi^2-1-2\del_1) + \frac{2\phi}{(\phi^2+\ga)^2} \left[ (\chi^2-1)^2 - \frac{\del_2}{4}(\chi-2)(\chi+1)^2 \right] &=& 0\\
\label{chieqn}
\chi'' - \frac{1}{(\phi^2+\ga)}(\chi^2-1)(4\chi-3\del_2/4) &=& 0
\eea
Given a static solution, the energy is
\beq
E[\phi,\chi]=\int dx\left(\frac{1}{2}\left(\phi'^2+\chi'^2\right) + V(\phi,\chi)\right).
\label{energy}
\eeq
We look for kinklike solutions interpolating between the false vacua $(\phi,\chi)=(\pm1,-1)$. We expect $\phi$ to be odd and $\chi$ even under space reflection, so we can solve the equations on the half-line $x\ge0$ with the following boundary conditions:
\bea
\phi(0) = 0,&\quad \chi'(0) = 0,&
\label{boundary1} \\
\phi(x)\rightarrow 1,&\quad \chi(x) \rightarrow -1&\quad {\rm as}\ \quad x \rightarrow \infty .
\label{boundary2}
\eea
The numerical approach used is an adaptation of the relaxation algorithm  explained beautifully in \cite{Press:1992}. Of course, numerically we do not integrate to infinity so \eqref{boundary2} must be handled differently. We integrate to some suitably large $x_{\rm max}$. The most obvious boundary condition would be to impose $(\phi,\chi)|_{x=x_{\rm max}}=(1,-1)$. However, this would produce a solution which diverges if extrapolated beyond $x_{\rm max}$. Instead, we linearize the equations (\ref{phieqn},\ref{chieqn}) about $(\phi,\chi)=(1,-1)$ and insist that the fluctuations tend to zero exponentially as $x\to\infty$. This gives linear relations between $(\phi,\phi')_{x=x_{\rm max}}$ and between $(\chi,\chi')_{x=x_{\rm max}}$ which we adopt as boundary conditions at $x_{\rm max}$. With these, the algorithm produces a solution which would extrapolate to $(\phi,\chi)=(1,-1)$ as $x\to\infty$.

\begin{figure}[hbt]
\begin{center}
\includegraphics[width=3.5in]{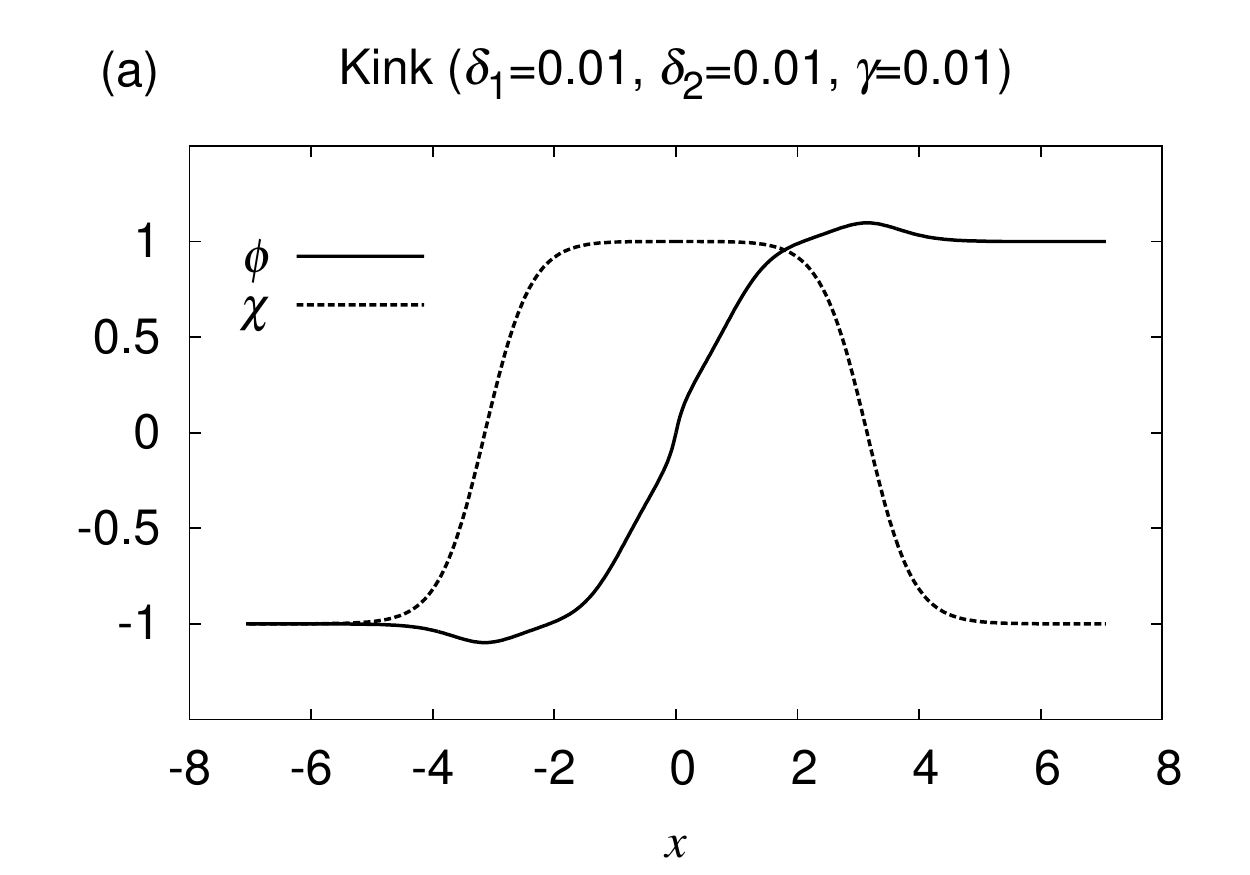}
\includegraphics[width=2.8in]{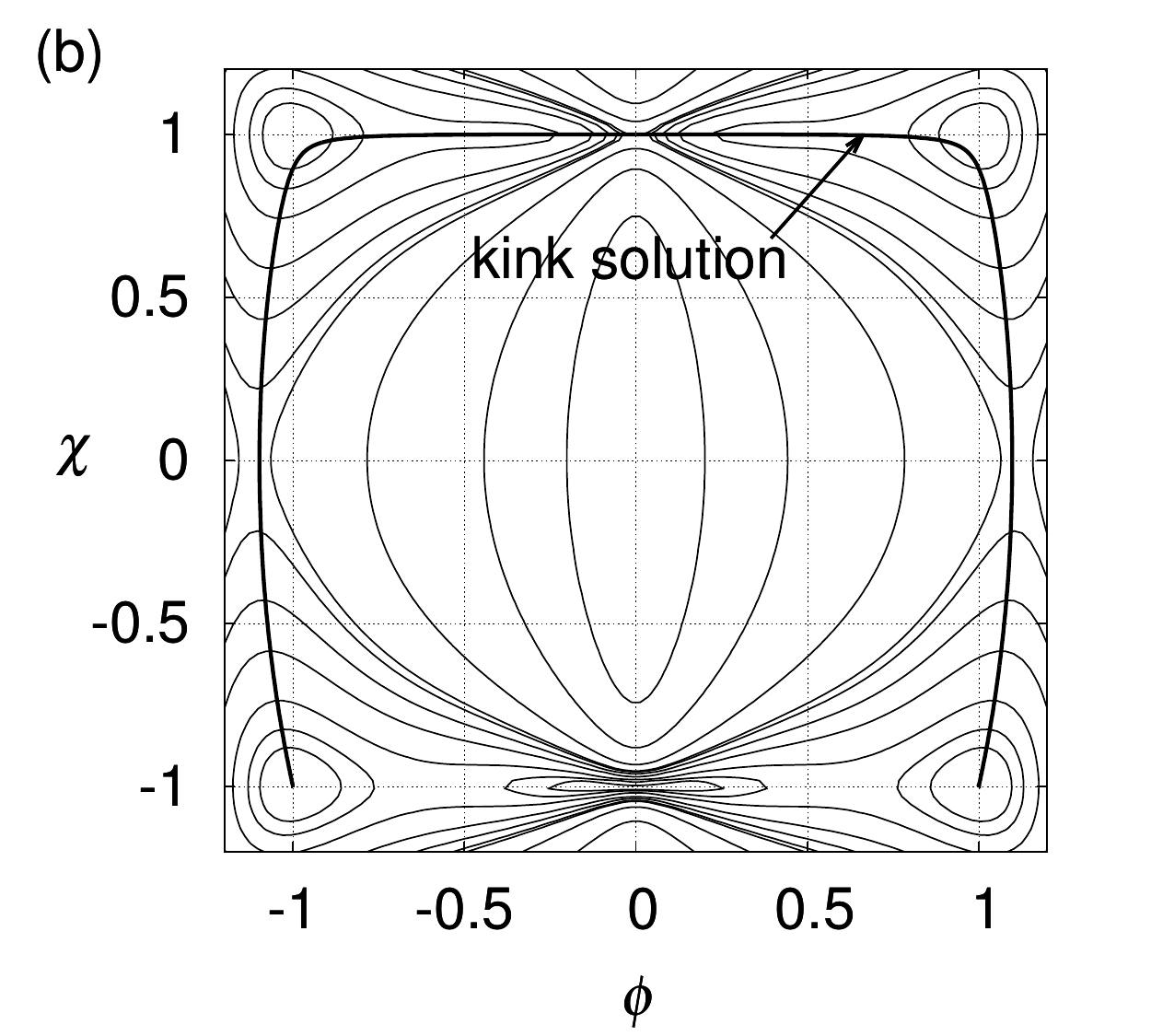}
\end{center}
\caption{(a) Graph and (b) parametric plot of solution for parameters $(\del_1,\del_2,\ga)=(0.01,0.01,0.01)$, looking much like that displayed in Fig.~\ref{fig4} if the widths of the regions (i) and (ii) depicted in that figure were collapsed to zero.
}
\label{fig5}
\end{figure}

Figure \ref{fig5} illustrates a typical solution, displaying the functions $\phi(x),\ \chi(x)$ in (a) and as a parametric plot in the $\phi\chi$ plane superimposed on a contour plot of the potential in (b). The latter is interesting because the equations of motion (\ref{phieqn},\ref{chieqn}) have a mechanical analogy: if $x$ is interpreted as a time coordinate and $(\phi,\chi)$ as Cartesian spatial coordinates, they are the equations of motion of a particle of unit mass moving in a potential $-V(\phi,\chi)$. Thus, for instance, the fact that $\phi$ goes beyond 1 at around $x=3$ (which at first sight might appear suspect) is actually perfectly reasonable, since the gradient of $V(\phi,\chi)$ along the positive $\phi$-axis points towards the origin beyond $x=1$. In other words, the particle of the mechanical analogy feels a force towards the origin, giving rise to the gently curved trajectory as the particle's position crosses $\chi=0$ (see Fig.~\ref{fig5}(b)). Also of interest, as we shall see, is the fact that in the centre of the soliton $\chi$ reaches $+1$ and is essentially constant while $\phi$ sweeps from near $-1$ to $+1$.

\begin{figure}[hbt]
\begin{center}
\includegraphics[width=3in]{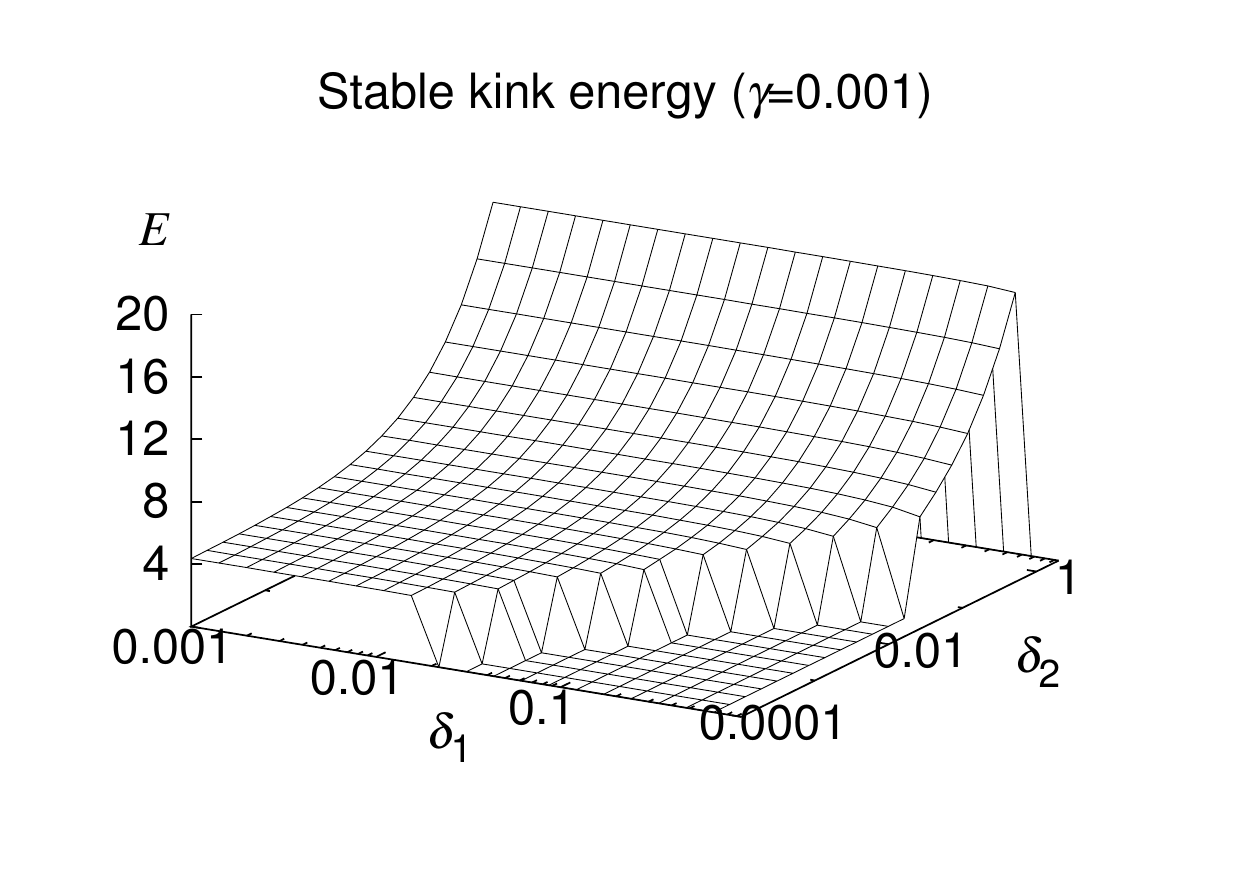}
\includegraphics[width=3in]{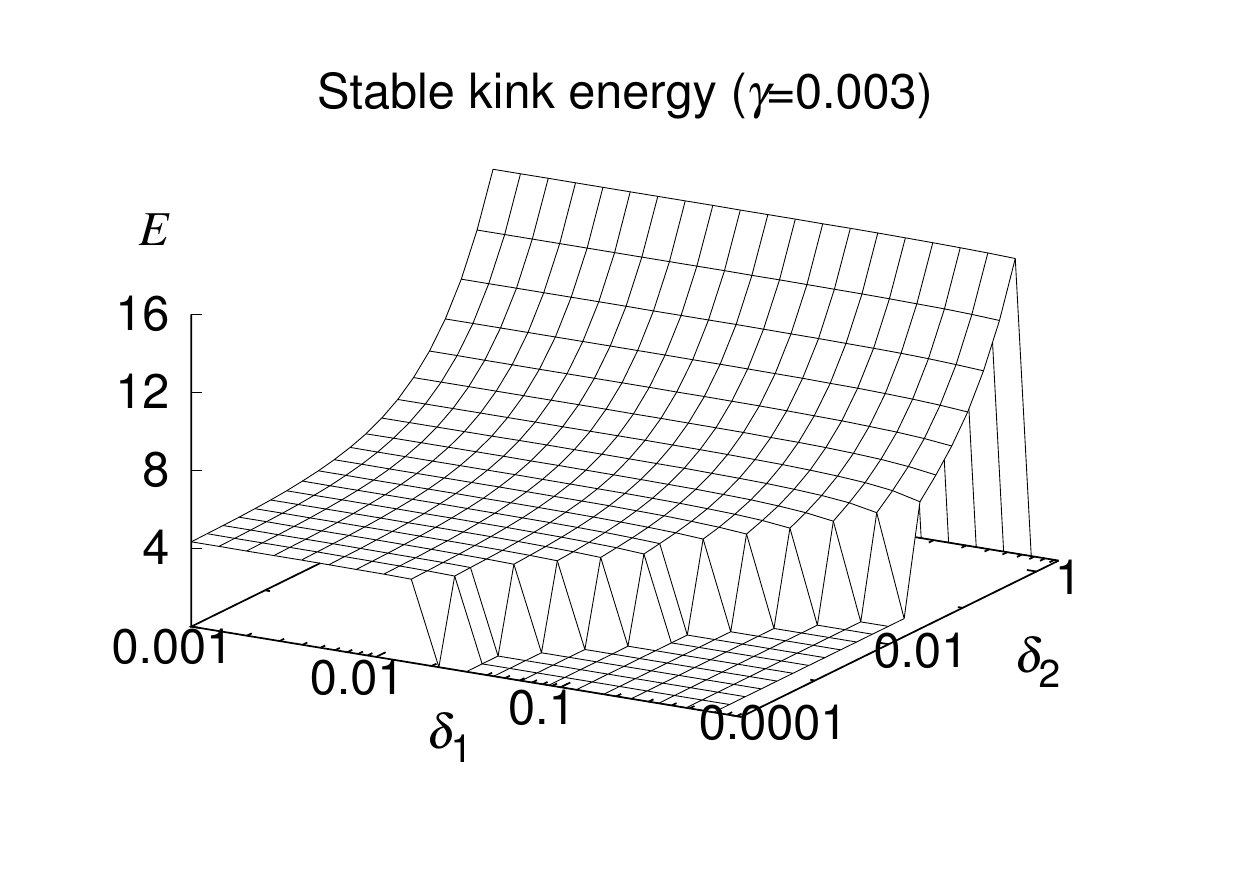}
\\
\includegraphics[width=3in]{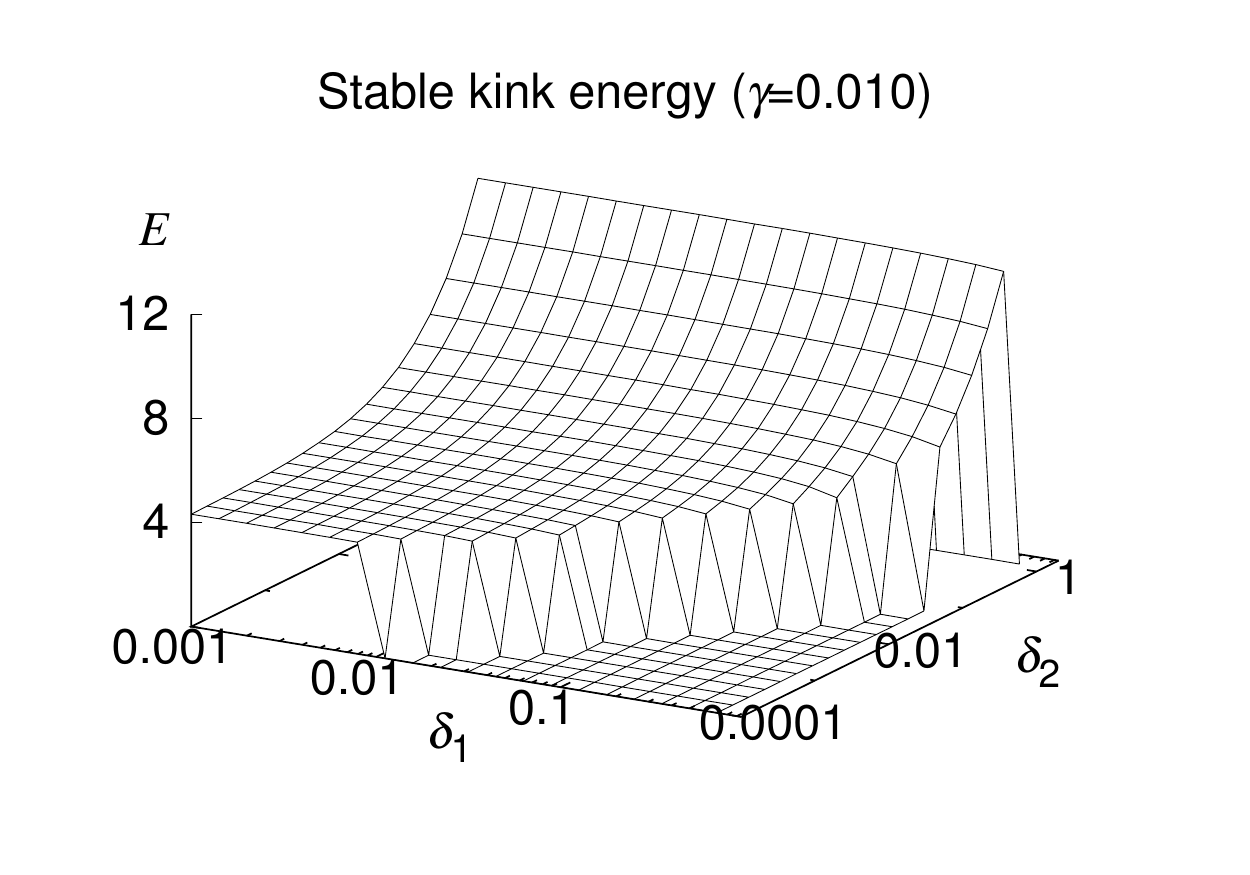}
\includegraphics[width=3in]{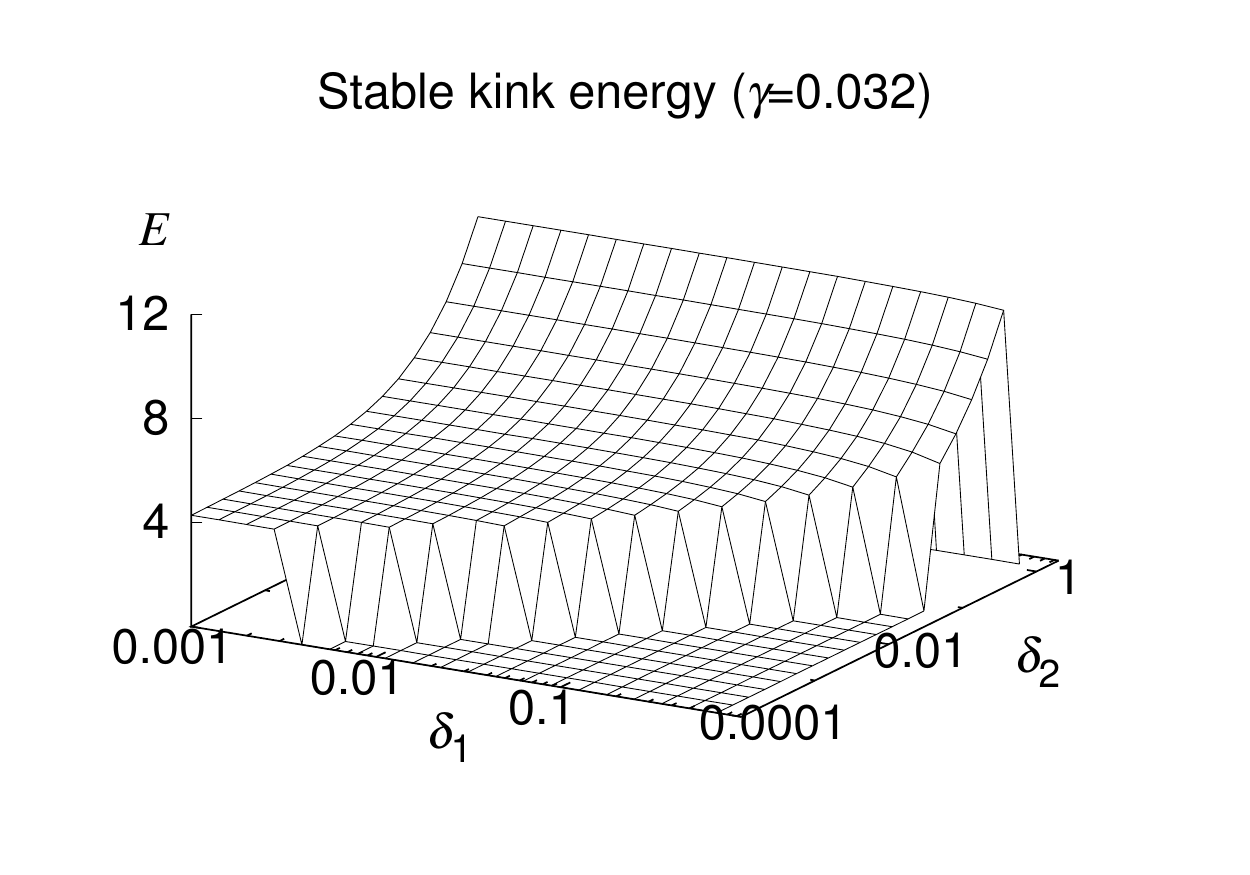}
\\
\includegraphics[width=3in]{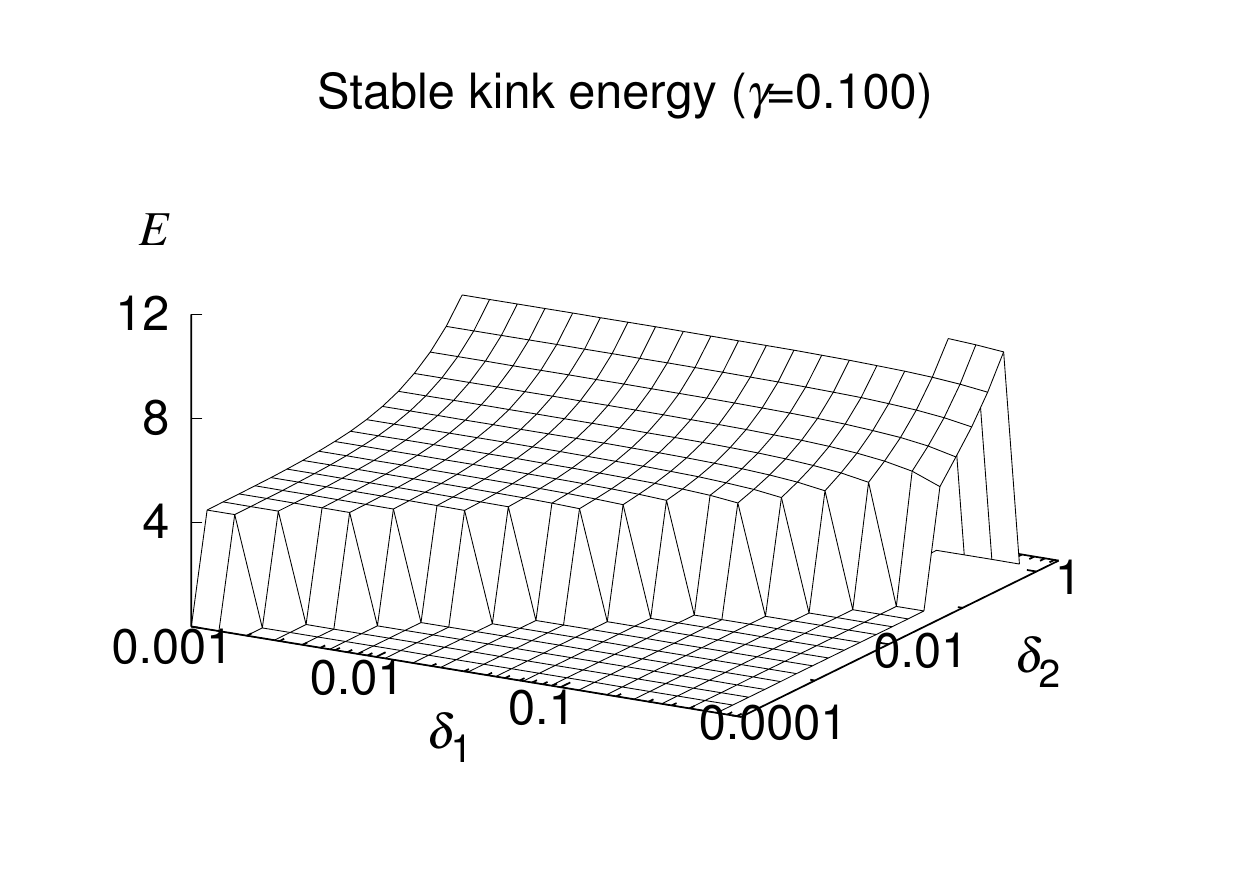}
\includegraphics[width=3in]{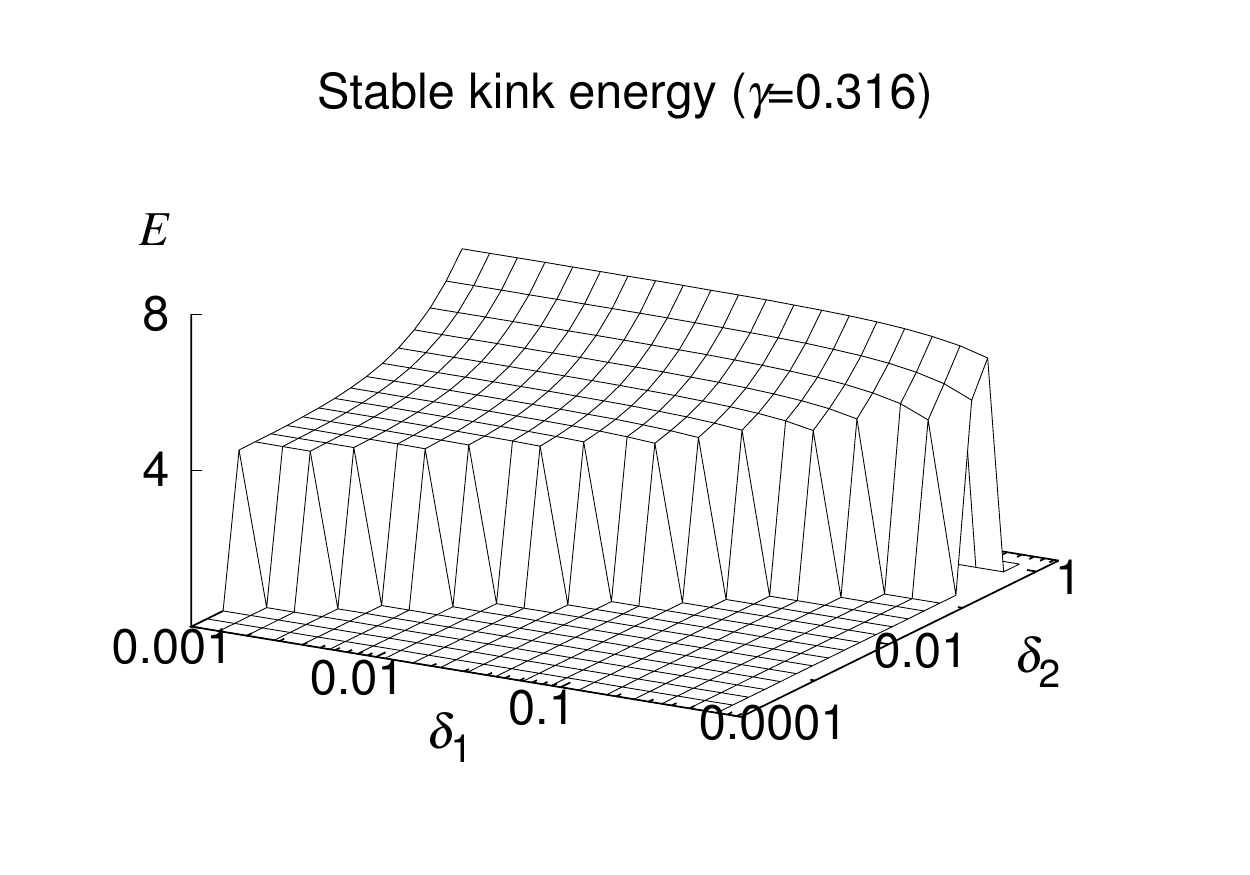}
\end{center}
\caption{Energy as a function of $\del_1,\del_2$ for six values of $\ga$. Where no solution was found (for example, the near corner of each plot), the energy was set to zero.}
\label{fig-solutions}
\end{figure}

A search for stable kinks was undertaken over a wide range of values of $\del_1,\del_2$ for six values of $\ga$. Where found, as a rule they look much like the one displayed in Fig.~\ref{fig5}.  The kink energy is displayed in Fig.~\ref{fig-solutions}. 
Although of course the details vary from one graph to the next, they share several striking features:
\begin{enumerate}
\item The energy is almost independent of $\del_1$, and increases as a function of $\del_2$.
\item For $\del_1$ large (approaching its maximum value of unity) and $\del_2\ll 1$, there is no solution; this region increases with $\ga$. (The nature of this stable/unstable transition will be explained below.)
\item For $\del_2$ somewhere in the neighbourhood of unity (the value depending on $\ga$ but very nearly independent of $\del_1$), there is no solution.
\end{enumerate}

\begin{figure}[hbt]
\begin{center}
\includegraphics[width=4in]{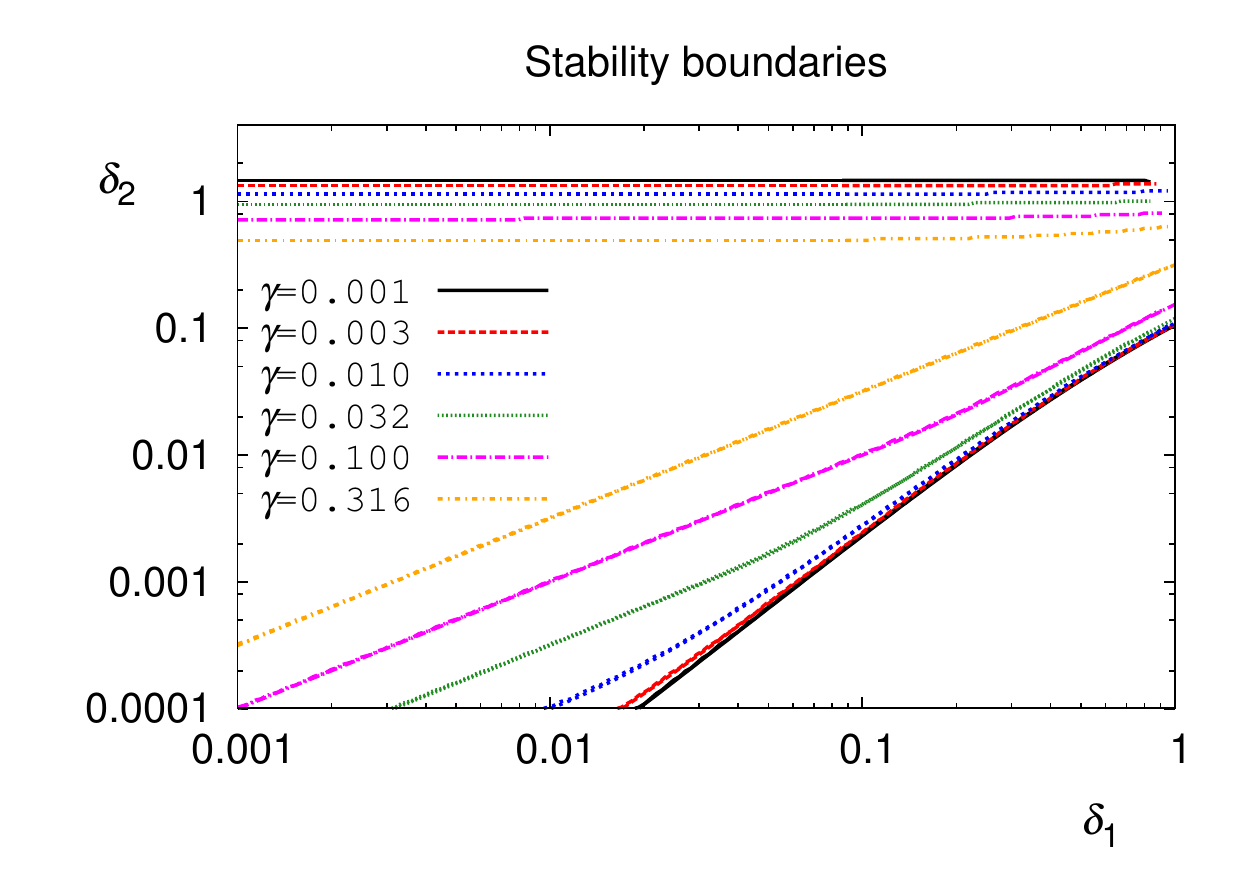}
\end{center}
\caption{(color online) Stability region (between the curves) in the $\del_1\del_2$ plane for each value of $\ga$ considered.}
\label{fig-boundaries}
\end{figure}

\begin{figure}[hbt]
\begin{center}
\includegraphics[width=3in]{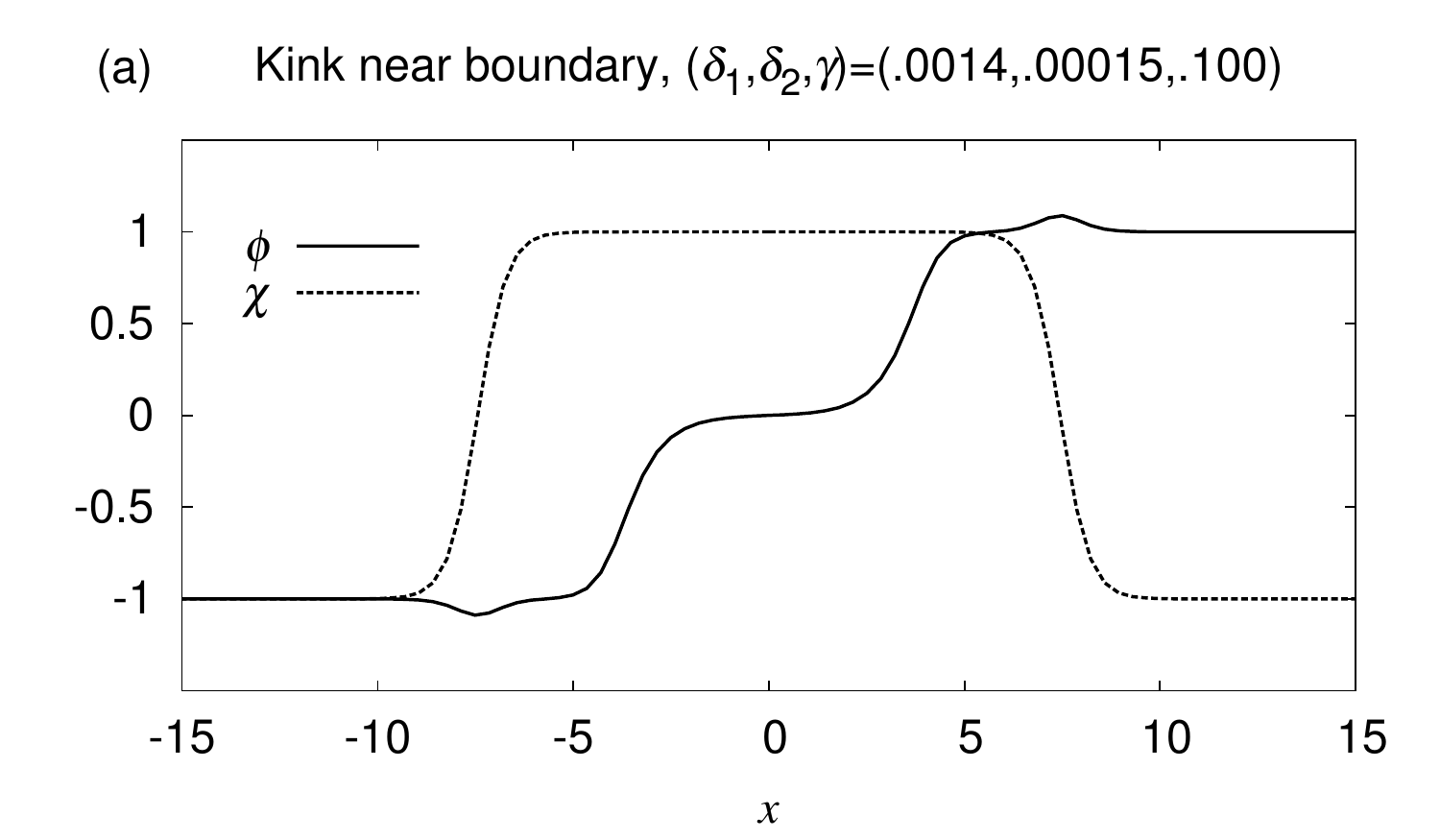}
\includegraphics[width=3in]{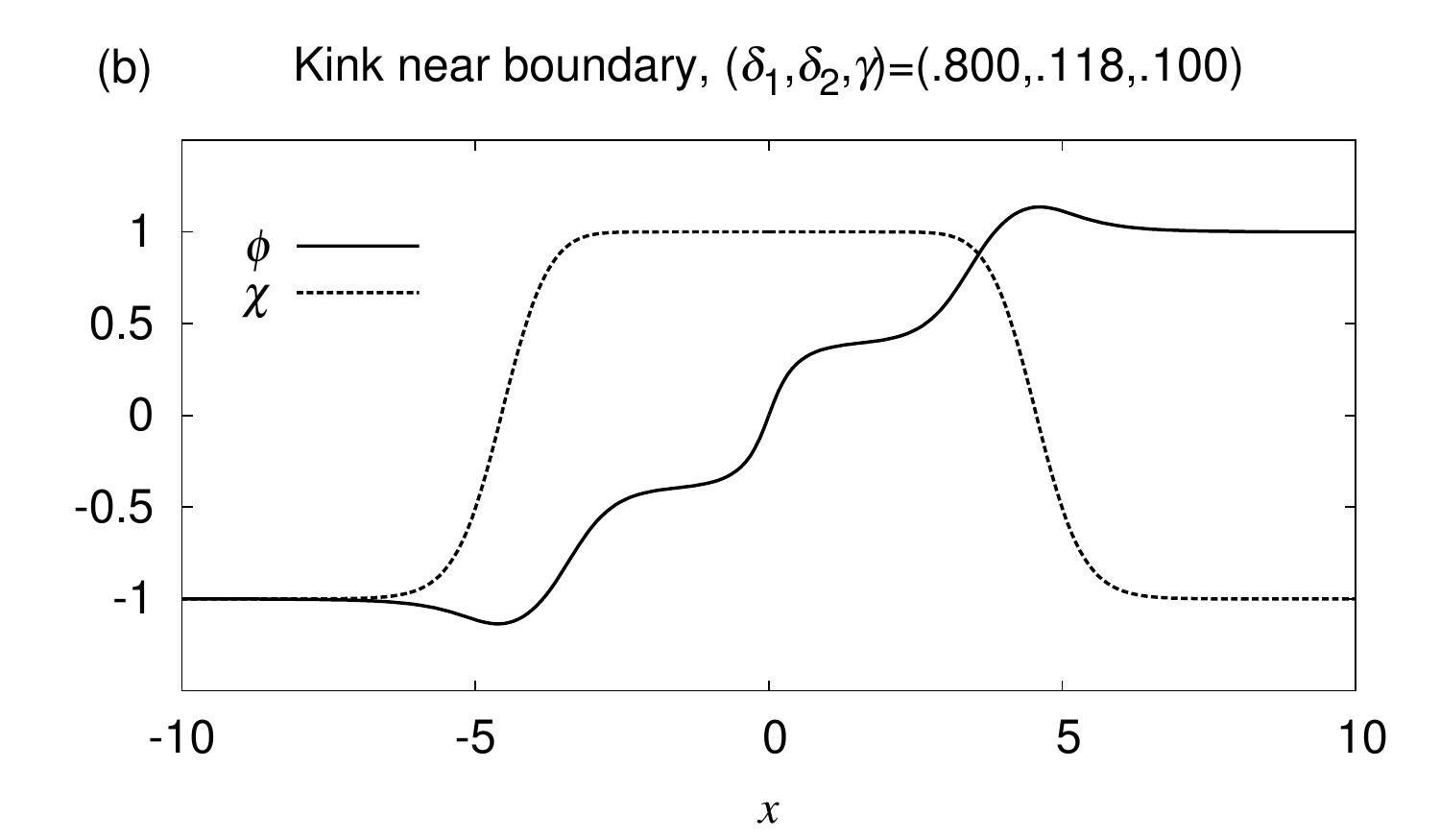}
\includegraphics[width=3in]{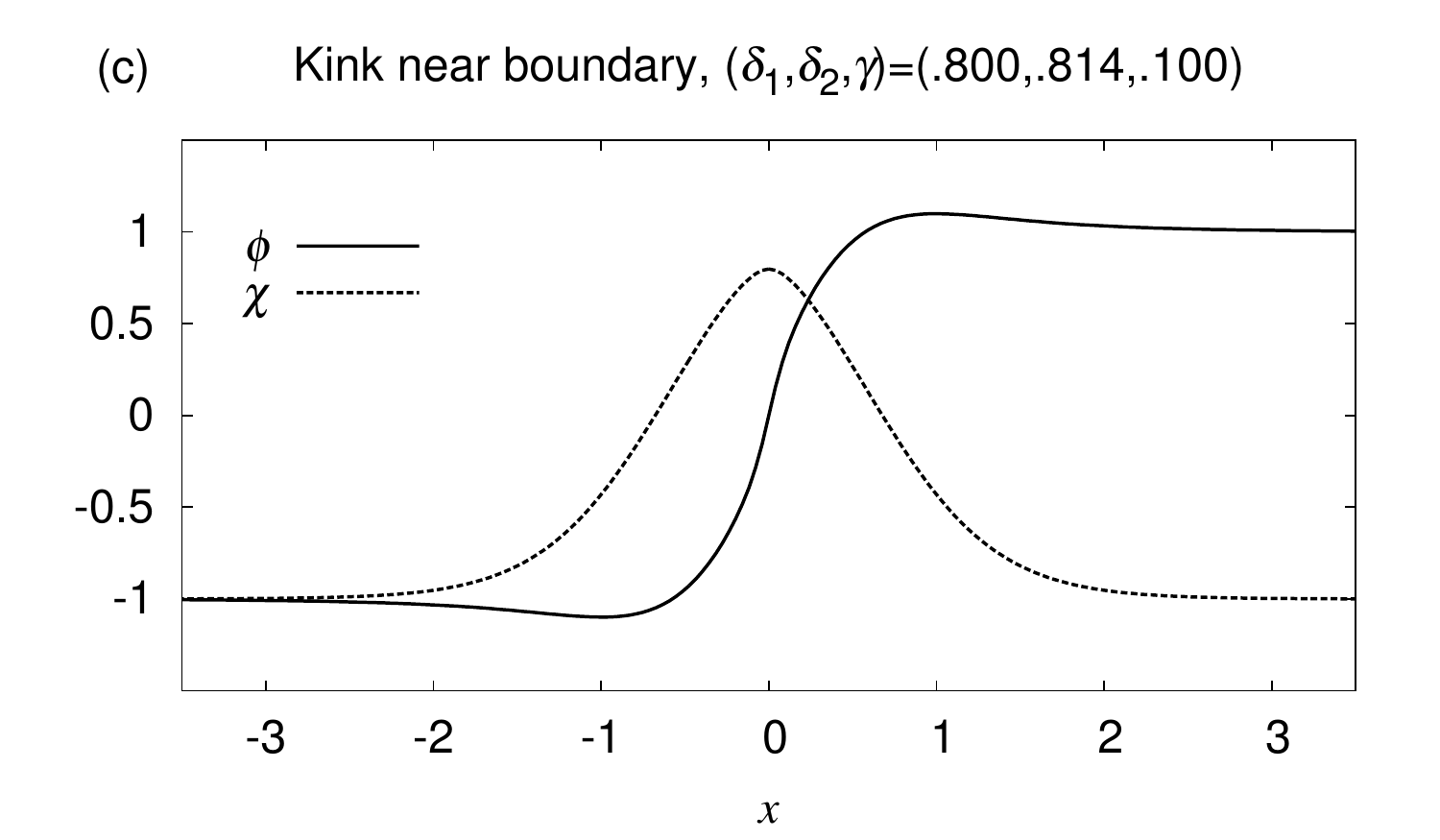}
\end{center}
\caption{Kinks near the stability boundary. (a) Lower boundary with inequality \eqref{cond1} satisfied. (b) Lower boundary with inequality \eqref{cond1} not satisfied. (c) Upper boundary.
}
\label{fig-knb}
\end{figure}

The latter two points are best illustrated in a plot of the stability regions in the $\del_1\del_2$ plane, for the various values of $\ga$, shown in Fig.~\ref{fig-boundaries}.
The six lower curves indicate an instability of the kink of the type discussed earlier (see the discussion around Eqs.~(\ref{cond1}-\ref{cond3})). As one approaches the stability boundary (imagine decreasing $\del_2$ from above), there is an easily-overlooked change of behaviour of $\phi$ at the centre of the soliton. Two possibilities are observed (see Fig.~\ref{fig-knb}(a,b)), depending on whether or not \eqref{cond1} is satisfied. If it is, the slope of $\phi$ decreases to zero at the origin, while if it is not, $\phi$ develops nonzero flat regions just off-centre (negative to the left, positive to the right). In both cases, $\phi$ is settling in towards the minimum (or the pair of minima) of $V(\phi,1)$. Below the stability line this minimum is of lower potential energy density than the false vacua at spatial infinity, and the kink becomes unstable, flying apart and leaving a lower-energy {\em false} vacuum (or pair of false vacua separated by a kink with $\chi=1$ and $\phi$ interpolating between the false vacua) in its wake. In fact, each of the lower lines in Fig.~\ref{fig-boundaries} is somewhat blurry because it is not one line but two that virtually coincide. One of these lines is the analytically-calculated stability line mentioned earlier (that is to say, \eqref{cond1} satisfied and \eqref{cond2} saturated, or alternatively \eqref{cond1} not satisfied and \eqref{cond3} saturated). The other line is the stability line found by the numerical scan of parameter space as described above. Obviously the excellent agreement between the two is a pleasing confirmation that our numerical work is behaving as expected.

The six upper curves indicate an instability of a very different type. As $\del_2$ increases, the potential $V_2(\chi)$ (see Fig.~\ref{fig2}) becomes more and more asymmetric and the barrier between $\chi=+1$ and $\chi=-1$ is reduced. Because of this, $\chi$ at the centre of the kink no longer reaches $+1$ (Fig.~\ref{fig-knb}(c)). The instability, then, is towards a configuration where $\chi$ no longer encircles the less-imposing potential barrier near $\phi=\chi=0$, after which it is energetically preferable for $\chi$ to reduce to $-1$ for all $x$, leaving the true vacuum in the centre of the soliton, which expands rapidly, converting false vacuum to true.

\section{Kink decay via tunneling \label{sec-tunneling}}
As we have seen, kinks are found over a wide range of parameters. They are classically stable, but since they are built out of the false vacuum, they (like the false vacuum itself) will decay via quantum tunneling. We will explore the decay of an isolated false kink in this section. In principle, we should solve the Euclidean field equations for the instanton (bounce) solution; the decay rate is expressible in terms of the action of the instanton, as we will see below.

However, this is a formidable task; even for ordinary vacuum decay the action has only been evaluated in a certain limit, where the false vacuum energy density is only slightly higher than that of the true vacuum. Then the so-called ``thin-wall approximation'' is valid, and the field theory problem is reduced to a single degree of freedom: the wall radius.

Here we will make a similar approximation, reducing the fields to a single degree of freedom which, at least near the stability boundary, should be a reasonable approximation to the field theory instanton. The latter has the lowest possible action, giving rise to the fastest possible decay rate; by making an approximation we will calculate an upper bound to the true instanton action which results in a lower bound to the decay rate of the kink.

Since we are interested in decay to the true vacuum, it is the second type of instability discussed in the previous section which is relevant, where the amplitude of the deviation of $\chi$ from its true vacuum value $\chi=-1$ goes to zero. As such, it is of interest to calculate the energy of a one-parameter family of static field configurations built out of the kink solution $(\phi_k(x),\chi_k(x))$, where the deformation parameter, written $h(t)$, modulates the amplitude of $\chi_k$ while not affecting $\phi_k$:
\[
\chi_k\to\chi_h(x,t) \equiv h(t)(\chi_k(x)+1)-1.
\]
$\chi_h$ interpolates between the true vacuum at $h=0$ and the kink at $h=1$. The energy of a static deformed configuration is $U(h) = E[\phi,\chi_h]$. Direct substitution and straightforward algebra yields
\beq
U(h) = \csta h^4 - \cstb h^3 + \cstc h^2 + \int dx \left(\frac{1}{2}\phi_k'^2 + V_1(\phi_k)\right)
\label{Uofh}
\eeq
where
\beq
\label{eq-abc}
\csta \equiv \int dx\, \frac{(\chi_k+1)^4}{\phi_k^2+\ga},\qquad
\cstb \equiv \left(4+\frac{\del_2}{4}\right)\int dx\, \frac{(\chi_k+1)^3}{\phi_k^2+\ga},\qquad
%\mbox{and}\qquad
\cstc \equiv \frac{3}{2}\cstb-2\csta.
\eeq
We note that $\csta$, $\cstb$ and $\cstc$ (which depend both explicitly and implicitly on $\del_1$, $\del_2$ and $\ga$) are positive; thus, the potential has a minimum at $h=1$, as indeed it must.

\begin{figure}[hbt]
\begin{center}
\includegraphics[width=3in]{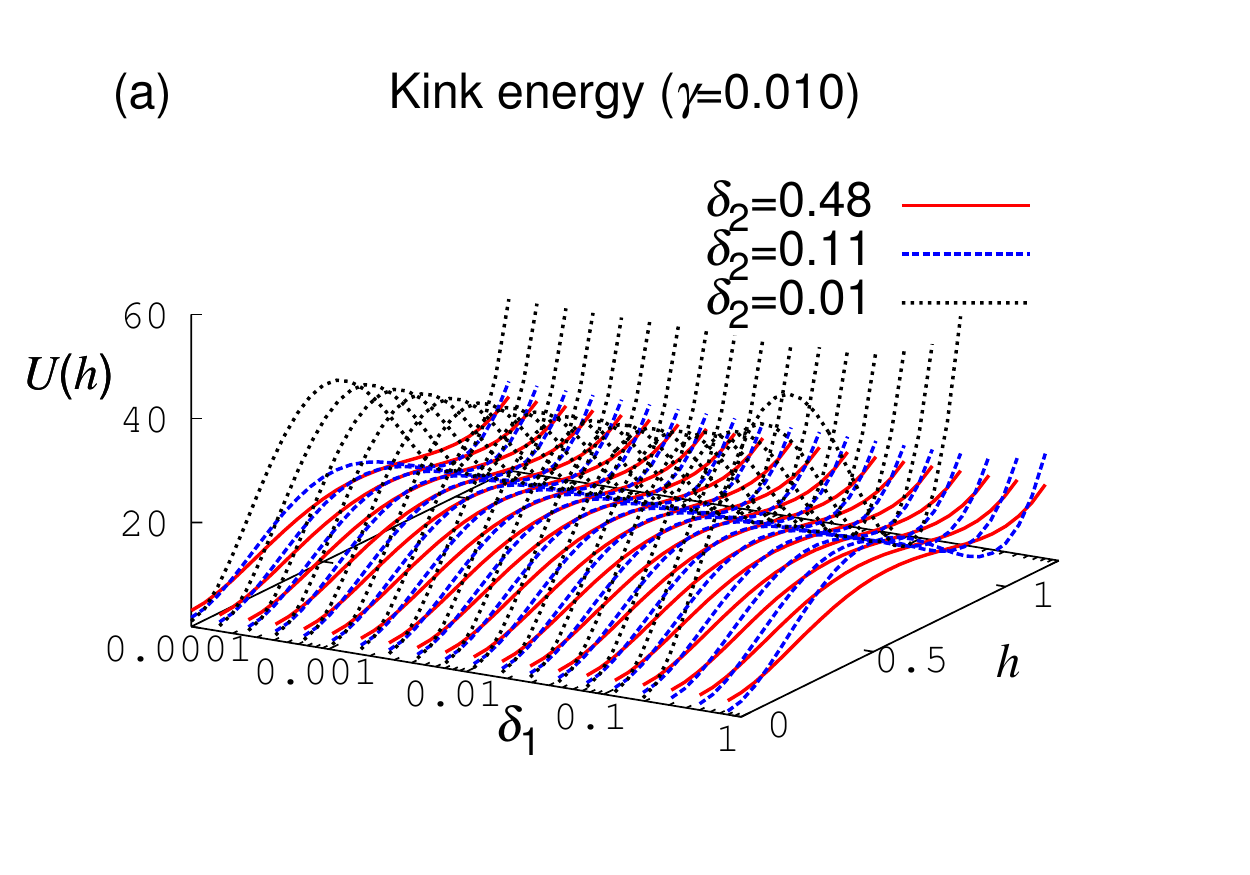}
\includegraphics[width=3in]{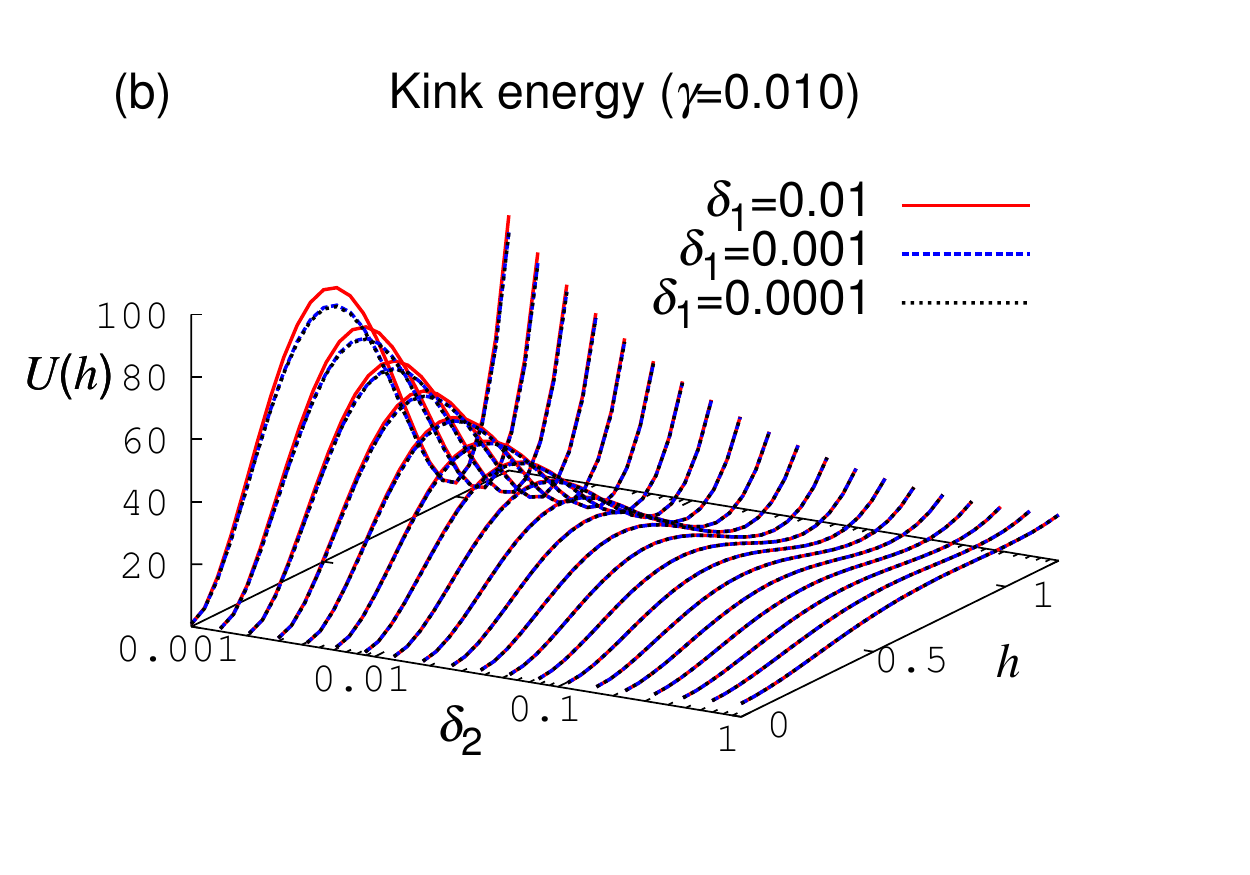}
\end{center}
\caption{(color online) Kink energy as a function of the deformation parameter $h$. 
}
\label{fig-U(h)}
\end{figure}

Fig.~\ref{fig-U(h)} shows $U(h)$, for $\ga=0.01$, displayed in two ways, highlighting the effect of $\del_1$ for three values of $\del_2$ in (a) and vice versa in (b). The relative unimportance of $\del_1$, already noted in Fig.~\ref{fig-solutions}, is readily seen in both of these. Changing $\ga$ produces only quantitative changes; thus for definiteness in the remainder of this section we will (unless stated otherwise) consider $\del_1=\ga=0.01$.

We see that in all cases there is an energy barrier separating the kink, at $h=1$, from a configuration of the same energy with $h$ near zero. This energy barrier must be overcome in a tunneling event. Clearly from Fig.~\ref{fig-U(h)}(b), the energy barrier decreases as $\del_2$ increases; this is as expected since we are approaching the second type of kink instability alluded to earlier. As $\del_2$ decreases, the energy barrier increases, which might be surprising given that here too we are approaching an instability. However, this instability is created by tunneling not to the true vacuum but rather to one of the false vacua  at $\chi=+1$ which arise for sufficiently small $\del_2$, in which case there is no reason to think that it will be particularly easy (in terms of the size of the energy barrier) to deform $\chi$ to the true vacuum.

\begin{figure}[hbt]
\begin{center}
\includegraphics[width=4in]{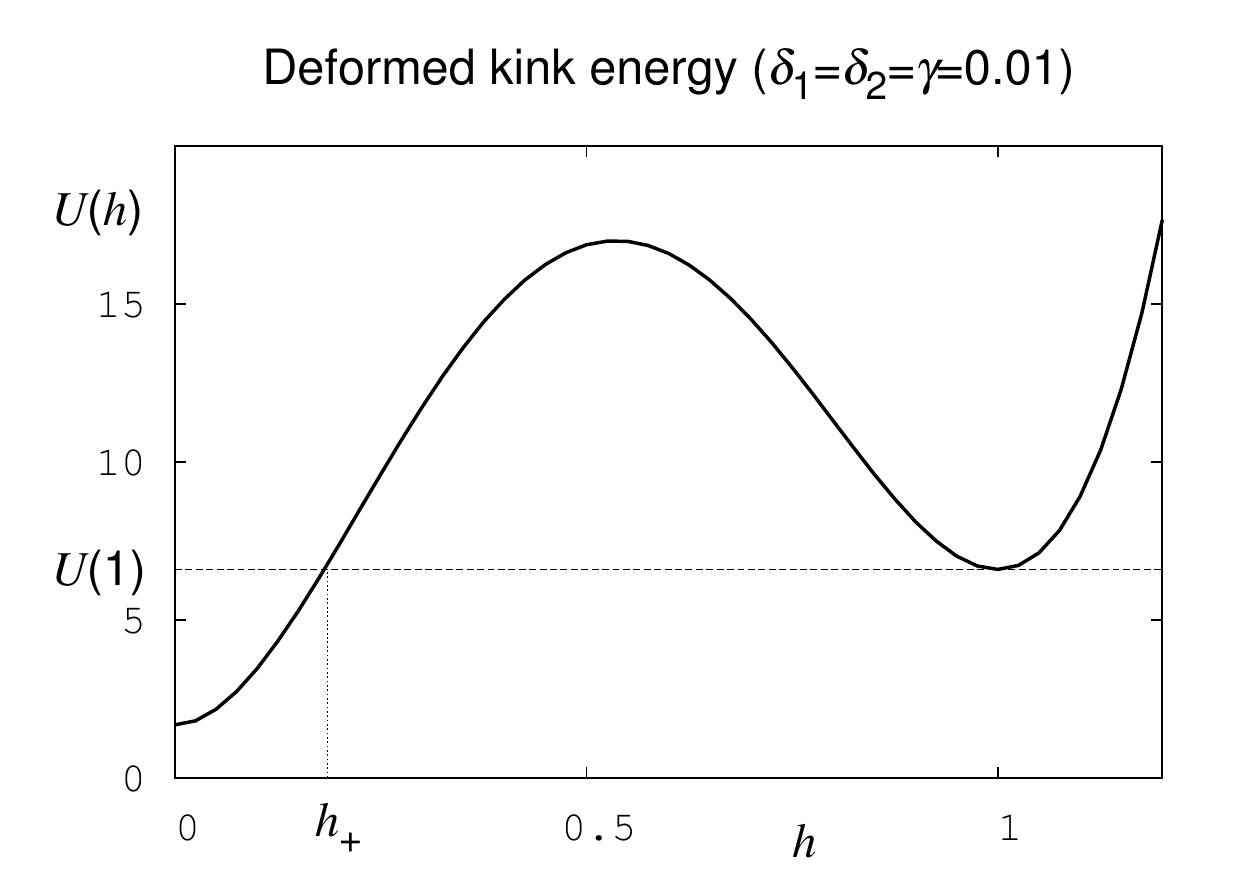}
\end{center}
\caption{Energy as function of deformation parameter $h$. 
}
\label{fig-Uofh-d1-01}
\end{figure}

The recipe for calculating the amplitude for quantum tunneling, which is related to the decay rate of the unstable state, is well-known \cite{Coleman:1977a}. For kink tunneling, having reduced the two fields to a single degree of freedom $h(t)$, the task is particularly straightforward. From the field theory Lagrangian density, we derive an effective Lagrangian for $h$ by the substitution
\[
\phi\to \phi_k, \qquad \chi\to \chi_{h(t)} = h(t)(\chi_k+1)-1.
\]
This yields
\[
L=\frac{1}{2}M {\dot h}^2 - U(h)
\]
where
\beq
\label{eq-def-m}
M = \int dx\, (\chi_k+1)^2
\eeq
and $U(h)$ is as above. Thus, the parameter $h$ can be thought of as the coordinate of a particle of mass $M$ in a potential $U(h)$. The tunneling amplitude is related to the Euclidean action of a particle moving in the potential $U_E(h)=-(U(h)-U(1))$. The relevant solution, the bounce, written $h_{\rm B}$, starts at $h=1$ at Euclidean time $\tau=-\infty$, rolls down towards the turnaround point $h_+$ (see Fig.~\ref{fig-Uofh-d1-01}) at $\tau=0$, and returns to $h=1$ as $\tau\to\infty$. Using standard methods, the action of the kink bounce is
\[
S_{\rm B,k} = \sqrt{8M}\int_{h_+}^1 dh \sqrt{U(h)-U(1)}.
\]

Now, $U(h)-U(1)$ is a quartic polynomial, and as can be seen from Fig.~\ref{fig-Uofh-d1-01} it has four real roots: the double root $h=1$, $h_+$, and another (negative) root which we will call $h_-$. Thus we can write 
\[
U(h)-U(1) = \csta(h-1)^2(h-h_+)(h-h_-);
\]
in terms of $\csta$ and $\cstb$ defined earlier,
\[
h_\pm = \frac{\cstb-2\csta \pm \sqrt{\cstb(\cstb-2\csta)}}{2\csta}.
\]
(Note that it is easy to show that $\cstb>2\csta$.) So the kink bounce action is
\beq
S_{\rm B,k} = \sqrt{8M\csta}\int_{h_+}^1 dh\,(1-h)\sqrt{(h-h_+)(h-h_-)}.
\label{SB1}
\eeq
The integral can be evaluated analytically, although the result is not terribly transparent; it is:
\bea
\label{eq-int1}
&&\frac{\sqrt{(1-h_+)(1-{h_-)}}}{24 }
\Bigg\{ 3 {h_+}^2-2 {h_+}{h_-} +3 {h_-}^2-4(h_+ + h_-) + 4\Bigg\}  \nonumber\\
&& +\frac{({h_+}-h_-)^2({h_+}+{h_-}-2)}{16}
\left( 2\log\left(\sqrt{1-{h_+}}+\sqrt{1-h_-}\right) - \log({h_+}-{h_-})\right),
\eea
which can also be written in terms of $\hat \cstb \equiv \cstb/\csta$:
\beq
\label{eq-int2}
\frac{1}{24}\sqrt{4-\frac{3}{2}\hat \cstb}\left( 3{\hat \cstb}^2 - 12\hat \cstb + 16 \right)
+\frac{\hat \cstb (\hat \cstb -2) (\hat \cstb -4)}{16}
\log\left( \frac{4 - \hat \cstb + 2\sqrt{4-\frac{3}{2}\hat \cstb}}{\sqrt{\hat \cstb (\hat \cstb -2)} }\right)
\eeq

\begin{figure}[hbt]
\begin{center}
\includegraphics[width=3.5in]{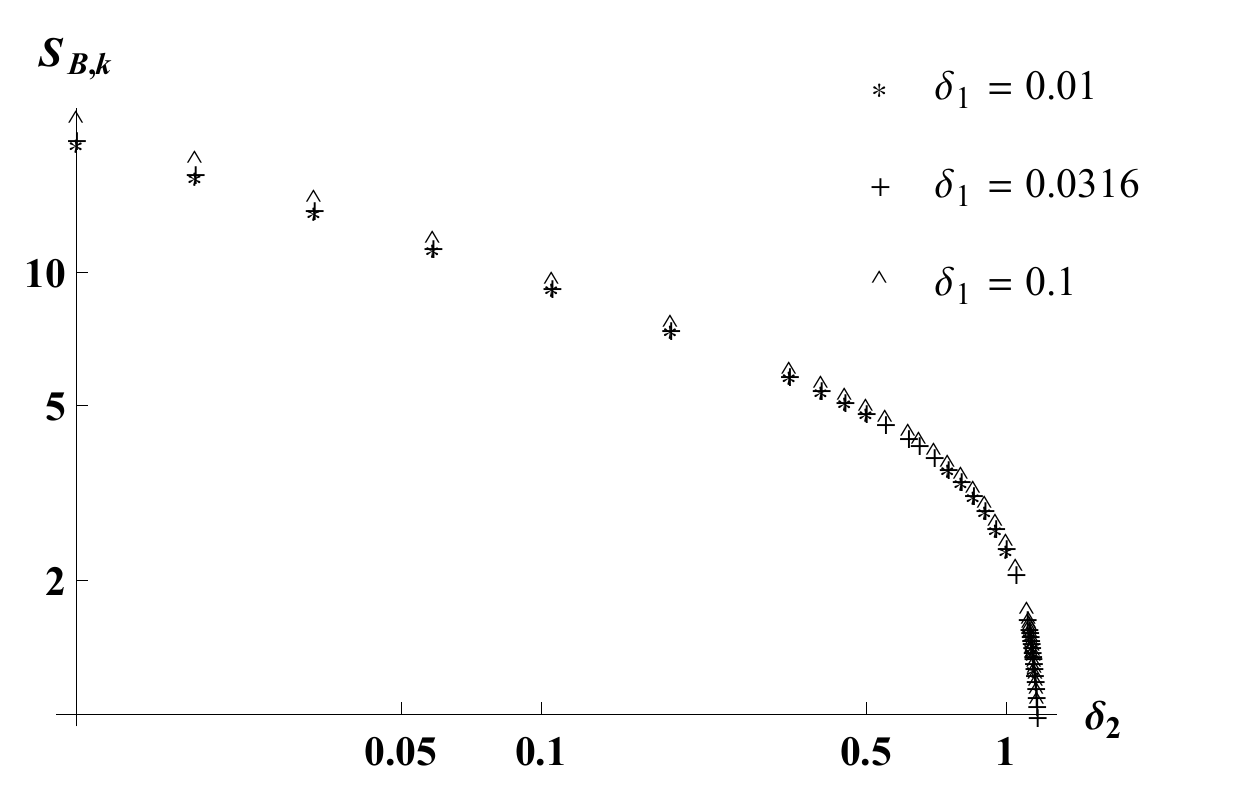}
\end{center}
\caption{Kink bounce action as a function of $\del_2$ for three values of $\del_1$; in all cases $\ga=0.01$.
}
\label{fig-kinkaction1}
\end{figure}

This expression depends on a very complicated way on the parameters. It is displayed in Fig.~\ref{fig-kinkaction1} as a function of $\del_2$ for three values of $\del_1$ and for $\ga=0.01$. The action goes to zero as $\del_2$ approaches the upper limit of the stability zone, as expected. Also as expected (see for example Fig. \ref{fig-U(h)}), the action is virtually independent of $\del_1$. 

%$%$ see below end doc

The kink decay rate is given by
\beq
\label{eq-kinkdecayrate}
\Gamma_{\rm k}= A_{\rm k} e^{-S_{\rm B,k}},
\eeq
where the prefactor $A_{\rm k}$ is fairly difficult to calculate and is beyond the scope of this paper. Nonetheless we will discuss it briefly in the next section.

\section{Kink-mediated vacuum decay? \label{sec-decay}}
In order to determine if and under what circumstances kinks have a significant effect on false vacuum decay, we must take into consideration three factors: the kink decay rate \eqref{eq-kinkdecayrate}, the kink density, and the ordinary vacuum decay rate. Let us examine the latter factor. We suppose that prior to tunneling the universe is in one of the false vacua $(\phi,\chi)=(\pm1,-1)$; for definiteness let us choose $\phi=+1$. The tunneling will reach the true vacuum $(\phi,\chi)=(0,-1)$. Since for any $\phi$ the potential $V(\phi,\chi)$ is minimized at $\chi=-1$ (see Fig.~\ref{fig2}), the bounce solution will be one of constant $\chi$. Thus the bounce is determined by a $\phi$-dependent Lagrangian obtained by substituting $\chi=-1$ in \eqref{eq-lagrangian1}, giving a rescaled version of \eqref{eq-lagrangian0} with potential depicted in Fig.~\ref{fig1}:
\[
\LL_\phi = \frac{1}{2}(\partial_\mu\phi)^2 - (\phi^2 - \del_1) (\phi^2 - 1)^2.
\]
Interestingly, the Lagrangian (and therefore any physical implications) depends only on $\del_1$.

\begin{figure}[hbt]
\begin{center}
\includegraphics[width=3.5in]{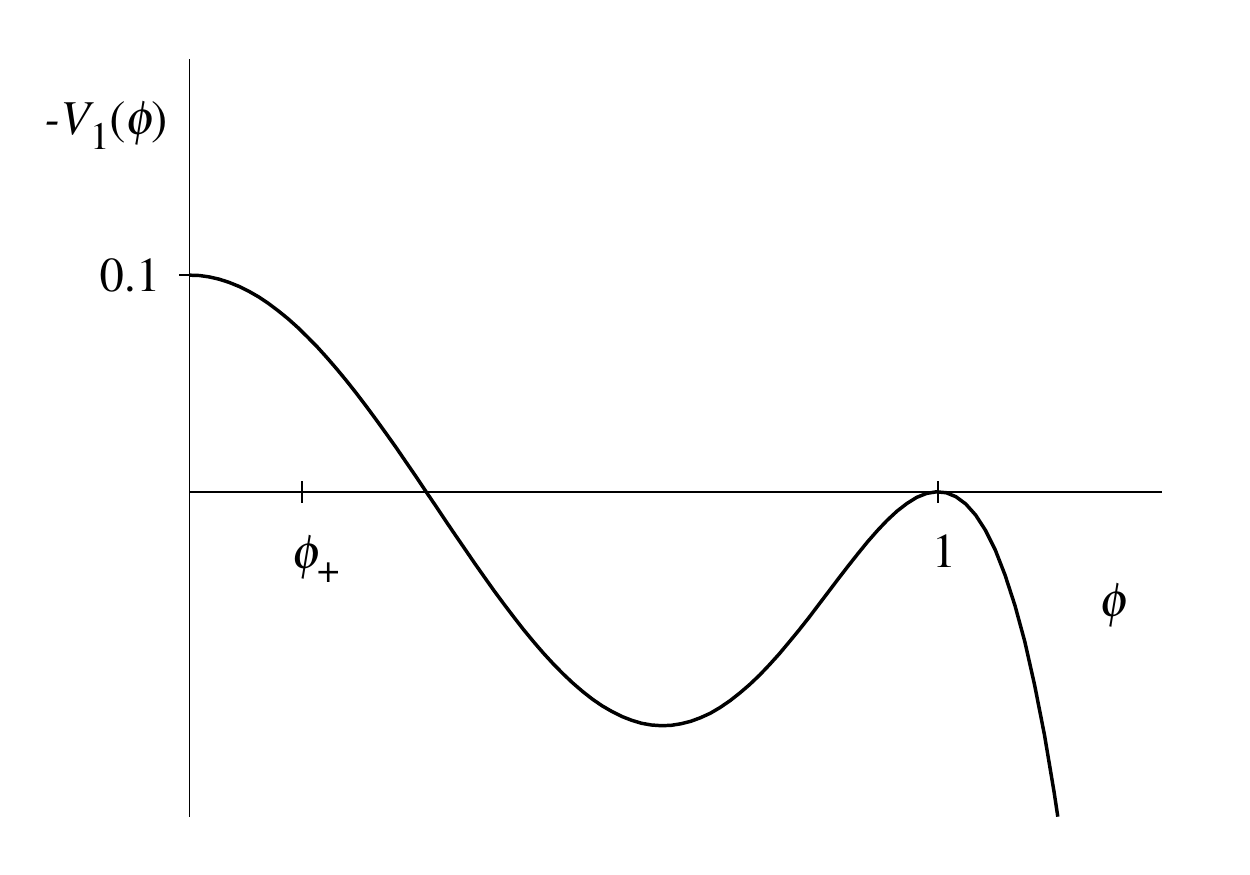}
\end{center}
\caption{Potential for a mechanical analogy helpful to understand the bounce ($\delta_1=0.1$). 
}
\label{fig-minusV1}
\end{figure}

\begin{figure}[hbt]
\begin{center}
\includegraphics[width=3.5in]{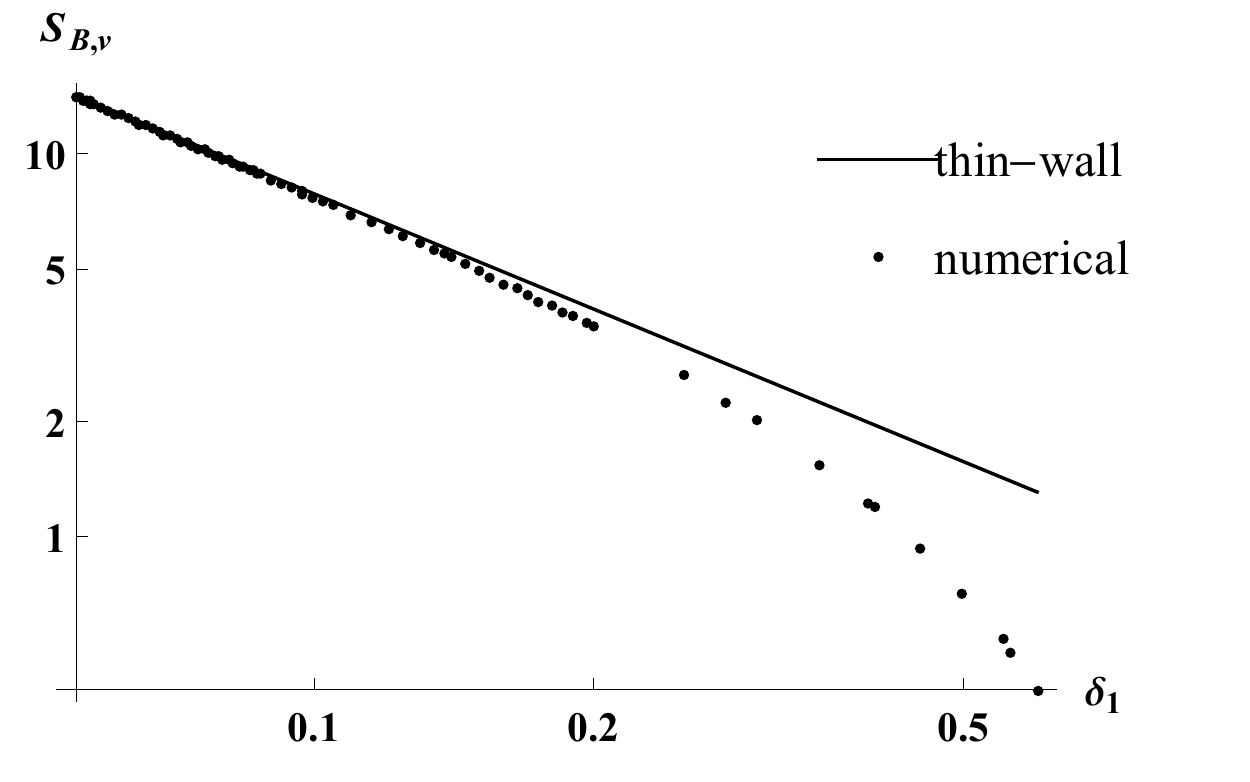}
\end{center}
\caption{Vacuum bounce action as a function of $\del_1$, calculated analytically using the thin-wall approximation (expected to be valid for $\del_1\ll1$) and numerically. The two agree in the domain of validity of the thin-wall approximation.}
\label{fig-actions1}
\end{figure}

As discussed in \cite{Coleman:1977a}, the bounce is the minimum-action solution of the Euclidean equation of motion (writing $\tau$ for the Euclidean time)
\[
(\partial_\tau^2 + \partial_x^2)\phi = \frac{d}{d\phi}V_1(\phi)
\]
with $\phi\to1$ at spacetime infinity. The minimum-energy solution is rotationally symmetric \cite{Coleman:1978}, so defining $r=\sqrt{x^2+\tau^2}$, the bounce solution based on the false vacuum $\phi_{\rm B,v}(r)$ is the solution of the following, where the prime denotes differentiation with respect to $r$:
\[
\phi'' + \frac{1}{r}\phi' = \frac{d}{d\phi}V_1(\phi),
\qquad \phi'(0)=0,\qquad \lim_{r\to\infty} \phi(r) =1.
\]
If we interpret $\phi$ as the position of a particle and $r$ as the time, this is the equation of motion of a particle moving in a potential $-V_1(\phi)$ (see Fig.~\ref{fig-minusV1}) with a time-dependent dissipative viscosity term; the particle starts at rest at a position to be determined and reaches $\phi=1$ at time infinity.
The initial point $\phi_+$ must have positive potential energy to compensate for the dissipation; a continuity argument \cite{Coleman:1977a} indicates that there is always a solution. The solution appears as a ``thin wall" if $\del_1\ll1$, in which case the solution and its action can be determined analytically; the latter is
\[
S_{\rm B,v} = \frac{\pi}{4\del_1}.
\]
The solution and action can be computed numerically for any $\del_1$; the analytical and numerical actions are displayed in Fig.~\ref{fig-actions1}. We see that the two agree for $\del_1\ll1$, as they must.

The vacuum decay rate per unit length is given by
\beq
\label{eq-vacdecayrate}
\Gamma_{\rm v}/L= A_{\rm v} e^{-S_{\rm B,v}},
\eeq
where again the prefactor is difficult to evaluate; we will discuss it briefly below.

Given the decay rate of a single kink \eqref{eq-kinkdecayrate} and the decay rate per unit length of the vacuum, we can now make a formal statement regarding the importance of kinks for vacuum stability. Imagine that the kink density (that is, the average number of kinks per unit length) is $\rho$. Then in a universe of length $L$ there are $\rho L$ kinks, and the decay rate (that is, the rate for {\em one} of the kinks to decay) is $\rho L\Gamma_{\rm k} = \rho LA_{\rm k} e^{-S_{\rm B,k}}$. We must compare this rate with that for a bubble to nucleate in the vacuum (that is, far from any kink). This is, assuming the total length occupied by kinks is much smaller than the total length, $(\Gamma_{\rm v}/L)L=A_{\rm v} e^{-S_{\rm B,v}}L$. We see there is a critical density of kinks where these two rates are equal:
\[
\rho_c = \frac{A_{\rm v}}{A_{\rm k}}e^{-(S_{\rm B,v}-S_{\rm B,k})}.
\]
This expression may look questionable because the density must have dimension $L^{-1}$, whereas the right hand side appears dimensionless. However, the factors $A_{\rm k},A_{\rm v}$ are proportional to ratios of square roots of determinants, and as we will now argue, they have different dimensions. In the case of $A_{\rm v}$, it is \cite{Callan:1977}
\beq
A_{\rm v} = \left(\frac{S_{\rm B,v}}{2\pi}\right)
\left|\frac{\det'\left(-\partial_\tau^2-\partial_x^2+V_1''(\phi_{\rm B,v})\right)}
{\det(-\partial_\tau^2-\partial_x^2+V_1''(\phi)|_{\phi=1})}\right|^{-1/2}
\label{eqn-Av}
\eeq
Let us focus on the determinant factors. These are determinants of the second variation of the Euclidean action about the solution (bounce in the numerator, false vacuum in the denominator). The denominator is straightforward since it is essentially the product of eigenvalues of a Klein-Gordon operator, whose eigenfunctions are plane waves. The numerator is more complicated since $\phi_{\rm B,v}$ depends on $x,\tau$. Thus the eigenfunctions are no longer plane waves. Furthermore, since the location of the bounce can be anywhere in the $x\tau$ plane, the operator has two zero modes: $\partial_x\phi_{\rm B,v}$ and $\partial_\tau\phi_{\rm B,v}$. The full determinant of this operator is therefore zero; however the prime in \eqref{eqn-Av} serves to remove these two zero eigenvalues from the determinant. Dimensionally, therefore, $A_{\rm v}$ has dimension (mass)$^{2}$.

In the case of $A_{\rm k}$, we have formally a similar expression:
\beq
A_{\rm k} = \left(\frac{S_{\rm B,k}}{2\pi}\right)^{1/2}
\left|\frac{\det'\left(-\partial_\tau^2+U''(h_{\rm B})\right)}
{\det(-\partial_\tau^2+U''(h)|_{h=1})}\right|^{-1/2}
\label{eqn-Ak}
\eeq
where $h_{\rm B}$ is the bounce configuration of the deformation variable $h$ as described in the previous section. One important difference is that there is now only one zero mode to remove in the determinant in the numerator (corresponding to time translation of the bounce); this is directly related to the change in power of the prefactor \cite{Callan:1977}. It also tells us that $A_{\rm k}$ has dimension (mass). 

There is no reason for the ratios of determinants to do anything pathological, so we can imagine that they are on the order of a fundamental mass scale of the problem (that of the scalar fields; call it $m$) to the appropriate power. Thus
\[
\rho_c\sim m\frac{S_{\rm B,v}}{\sqrt{2\pi S_{\rm B,k}}}e^{S_{\rm B,k}-S_{\rm B,v}}.
\]
It is clear that as the parameters -- and therefore the actions -- vary, the dominant effect on $\rho_c$ is due to the exponential factor (indeed, this is why virtually all discussions of tunneling ignore the prefactor). Fig.~\ref{fig-kinkaction1} tells us that $S_{\rm B,k}$ is essentially independent of $\del_1$ over a wide range, while it depends substantially on $\del_2$. In contrast, $S_{\rm B,v}$ (Fig.~\ref{fig-actions1}) depends only on $\del_1$. Examining these figures in detail, we see that if $\del_1$ is small (much less than 1, say) while $\del_2$ is of order 1, the exponential factor will be miniscule and even a very dilute presence of kinks will have a dramatic effect on vacuum decay.

\section{Conclusions \label{sec-conclusions}}
We have examined the effect of topological solitons on vacuum decay in a toy model in 1+1 dimensions with symmetry-breaking false vacua and a symmetry-restoring true vacuum. It is far from automatic that such a model would have solitons; for instance, in the simplest such model any configuration interpolating between the false vacua would be unstable, with two ``half-solitons" repelling each other and leaving true vacuum in their wake. The model we study is not particularly realistic but it does show the existence of models with classically stable solitons.

We find that solitons can indeed have an important effect on vacuum stability. This is essentially because the energy barrier which makes the soliton classically stable, on the one hand, and that which makes the vacuum itself classically stable, on the other, are independent. In particular, while maintaining a large barrier between true and false vacua (resulting in a very long-lived vacuum), the barrier stabilizing the kink can be made small. In this case, the presence of even a very dilute gas of kinks would cause the vacuum to decay rapidly via nucleation of a bubble of true vacuum in the vicinity of one of the kinks.

Of course, any model with solitons in 1+1 dimensions will have domain walls in 3+1 dimensions. The decay through tunneling of a domain wall in 3+1 dimensions would be very different than that of the corresponding kink in 1+1 dimensions: the wall would develop a bulge somewhere and the bulge would then expand, analogous to the corresponding phenomenon with stringlike solitons \cite{Lee:2013b}. This situation is currently under investigation.

\section*{Acknowledgements}
This work was financially supported in part by the Natural Science and Engineering Research Council of Canada.

%\newpage

\bibliographystyle{unsrt}

\bibliography{mybibdatabase}

\end{document}